\pdfoutput=1
\documentclass[aip,amsmath,amssymb, reprint]{revtex4-1}

\usepackage{graphicx}
\usepackage{dcolumn}
\usepackage{bm}
\usepackage{booktabs}
\usepackage{multirow}
\usepackage[utf8]{inputenc}
\usepackage[T1]{fontenc}
\usepackage[version=3]{mhchem}
\usepackage{hyperref} 
\usepackage{rotating} 

\usepackage{notes2bib}

\newcommand{\revision}[1]{#1}

\newcommand*{\citeref}[1]{Ref.~\citenum{#1}}
\newcommand*{\citerefs}[1]{Refs.~\citenum{#1}}

\newcommand*{\CTv}{\ensuremath{\text{C}_\text{2v}}}
\newcommand*{\DTh}{\ensuremath{\text{D}_\text{2h}}}
\newcommand*{\DSh}{\ensuremath{\text{D}_\text{6h}}}
\newcommand*{\Cs}{\ensuremath{\text{C}_\text{s}}}

\let\eqref\undefined
\newcommand*{\eqref}[1]{Eq.~(\ref{eq:#1})}
\newcommand*{\tabref}[1]{Table~\ref{tab:#1}}

\newcommand*{\figref}[1]{Fig.~\ref{fig:#1}}

\newcommand*{\secref}[1]{Section~\ref{sec:#1}}

\usepackage{braket}
\usepackage[table]{xcolor}
\usepackage{makecell}
\usepackage{listings}
\usepackage{bbold}
\usepackage{url}
\usepackage{mathtools}

\usepackage{colortbl}

\usepackage{placeins}
\usepackage{subfigure}

\newcommand{\PySCF}{\textsc{PySCF}}
\newcommand{\PyFLOSIC}{\textsc{PyFLOSIC}}
\newcommand{\ERKALE}{\textsc{ERKALE}}

\newcommand{\fodMC}{\textsc{fodMC}}

\definecolor{darkolivegreen}{rgb}{0.33, 0.42, 0.18}
\definecolor{darkpink}{rgb}{0.91, 0.33, 0.5}
\definecolor{aqua}{rgb}{0.0, 1.0, 1.0}
\definecolor{applegreen}{rgb}{0.55, 0.71, 0.0}
\definecolor{blizzardblue}{rgb}{0.67, 0.9, 0.93}
\definecolor{bleudefrance}{rgb}{0.19, 0.55, 0.91}
\definecolor{desertsand}{rgb}{0.93, 0.79, 0.69}
\definecolor{mediumspringbud}{rgb}{0.79, 0.86, 0.54}
\definecolor{rosequartz}{rgb}{0.67, 0.6, 0.66}

\definecolor{armygreen}{rgb}{0.29, 0.33, 0.13}
\definecolor{asparagus}{rgb}{0.53, 0.66, 0.42}
\definecolor{aquamarine}{rgb}{0.5, 1.0, 0.83}
\definecolor{cambridgeblue}{rgb}{0.64, 0.76, 0.68}

\definecolor{background-color}{gray}{0.98}

\usepackage[color=red]{attachfile2}
\usepackage{soul}

\begin{document}

\title{Chemical bonding theories as guides for self-interaction corrected solutions: multiple local minima and symmetry breaking}
\author{Kai Trepte}
\email{ktrepte@stanford.edu/kai.trepte1987@gmail.com}
\affiliation{\mbox{Stanford University, SUNCAT Center for Interface Science and Catalysis, Menlo Park, CA 94025, USA}}
\author{Sebastian Schwalbe}
\email{schwalbe@physik.tu-freiberg.de}
\affiliation{\mbox{Institute of Theoretical Physics, TU Bergakademie Freiberg, D-09599 Freiberg, Germany}}
\author{Simon Liebing}
\affiliation{\mbox{Joint Institute for Nuclear Research Dubna, Bogoliubov Laboratory of Theoretical Physics, 141980 Dubna, Russia}}
\author{Wanja T. Schulze}
\affiliation{\mbox{Institute of Theoretical Physics, TU Bergakademie Freiberg, D-09599 Freiberg, Germany}}
\author{Jens Kortus}
\affiliation{\mbox{Institute of Theoretical Physics, TU Bergakademie Freiberg, D-09599 Freiberg, Germany}}
\author{Hemanadhan Myneni}
\affiliation{\mbox{Science Institute and Faculty of Physical Sciences, University of Iceland  VR-III, 107 Reykjav\'{i}k, Iceland}}
\author{Aleksei V. Ivanov}
\affiliation{\mbox{Science Institute and Faculty of Physical Sciences, University of Iceland  VR-III, 107 Reykjav\'{i}k, Iceland}}
\author{Susi Lehtola}
\email{susi.lehtola@alumni.helsinki.fi}
\affiliation{Molecular Sciences Software Institute, Blacksburg, VA 24061, USA}

\date{\today}

\begin{abstract} 
Fermi--L\"owdin orbitals (FLO) are a special set of localized orbitals, which have become commonly used in combination with the Perdew--Zunger self-interaction correction (SIC) in the FLO-SIC method.
The FLOs are obtained for a set of occupied orbitals by specifying a classical position for each electron.
These positions are known as Fermi-orbital descriptors (FODs), and they have a clear relation to chemical bonding.
In this study, we show how FLOs and FODs can be used to initialize, interpret and justify SIC solutions in a common chemical picture, both within FLO-SIC and in traditional variational SIC, and to locate distinct local minima in either of these approaches.

We demonstrate that FLOs based on Lewis' theory lead to symmetry breaking for benzene---the electron density is found to break symmetry already at the symmetric molecular structure---while ones from Linnett's double-quartet theory reproduce symmetric electron densities and molecular geometries.
Introducing a benchmark set of 16 planar, cyclic molecules, we show that using Lewis' theory as the starting point can lead to artifactual dipole moments of up to 1 Debye, while Linnett SIC dipole moments are in better agreement with experimental values.
We suggest using the dipole moment as a diagnostic of symmetry breaking in SIC and monitoring it in all SIC calculations.
We show that Linnett structures can often be seen as superpositions of Lewis structures and propose Linnett structures as a simple way to describe aromatic systems in SIC with reduced symmetry breaking.
The role of hovering FODs is also briefly discussed.
\end{abstract}

\maketitle

\section{Introduction}
\label{sec:intro}
One of the simplest descriptions of chemistry is given by Lewis' theory (LT) of chemical bonding.\cite{Lewis1916_762} 
The main focus of LT is the octet rule, which states that the valence electrons of main group elements are arranged in four pairs that imitate the electron configuration of noble gas atoms.\cite{Lewis1916_762} 
While LT is probably the best-known, most extensively taught, and most widely used theory of chemical bonding, there are several cases in which LT does not suffice for an accurate understanding of chemical bonding.\cite{Liu2020_1} 
For instance, as LT assumes that the electrons are always paired---thereby forming a closed-shell singlet state---LT is not able to describe molecules with non-singlet ground states, such as the oxygen molecule.
This shortcoming of LT, the lack of electronic spin, is solved by Linnett's  double-quartet (LDQ) theory\cite{Linnett1960_859, Linnett1961_2643, Luder1964_55, Empedocles1964_166, Linnett1964_1} that explicitly includes the electronic spin in the formalism. 
The nearly-forgotten\cite{Jensen2017_74} LDQ replaces the octet of LT by two quartets---one quartet per spin channel---hence the name of the theory.

\citet{Schwalbe2019_2843} revived LDQ theory in the context of the Perdew--Zunger (PZ) self-interaction correction\cite{Perdew1981_5048} (SIC) to density functional theory (DFT).\cite{Hohenberg1964_B864, Kohn1965_A1133}
The idea of SIC is to approximately remove one-electron self-interaction error from the Kohn--Sham (KS) energy functional\cite{Kohn1965_A1133} $E^\text{KS}$
\revision{
\begin{align}
 \nonumber   E^{\text{KS}}[n^\alpha,n^\beta] = & \ T_{\text{s}}[n^\alpha,n^\beta]+ V[n] \ \\ & + E_J[n] + E_\text{xc}[n^\alpha,n^\beta]\;, \label{eq:ksE}
\end{align}
where $T_{\text{s}}[n^\alpha,n^\beta]$ is the kinetic energy of the non-interacting system, $V[n]$ is the external potential energy, and $n^\sigma$ is the electron density for spin $\sigma$, with $\alpha$ and $\beta$ representing spin up and spin down, respectively. The PZ functional explicitly removes the self-interaction error one electron at a time}\cite{Perdew1981_5048}
\begin{equation}
    E^\text{PZ} = E^\text{KS} - \sum_{i\sigma} \left( E_J[n_{i\sigma}]+E_\text{xc}[n_{i\sigma}] \right), \label{eq:pz}
\end{equation}
where $n_{i\sigma}$ is the electron density of the $i$-th occupied orbital with spin $\sigma$, and $E_J$ and $E_\text{xc}$ denote the Coulomb and exchange-correlation energy functionals.

While the usual formulation of DFT is invariant to rotations between occupied orbitals,\cite{Lehtola2020_1218} this invariance is broken by SIC which explicitly depends on the occupied orbitals used to evaluate \eqref{pz}.\cite{Pederson1984_1972}
The PZ functional then needs to be minimized in terms of the $\mathcal{O}(N_\text{occ}^2)$ rotation angles for $N_\text{occ}$ occupied spatial orbitals; the orbitals tend to become localized upon the optimization.\cite{Pederson1984_1972}
\revision{It has been shown more recently that the variational optimization requires complex orbitals;\cite{Lehtola2016_3195} software for such calculations is openly available.\cite{Lehtola2014_5324, Lehtola2016_3195}}

In the Fermi--Löwdin orbital\cite{Luken1982_265} (FLO) formulation of SIC,\cite{Pederson2014_121103, Pederson2015_064112, Pederson2015_153, Yang2017_052505} the need for occupied-occupied orbital rotations is avoided by the use of Fermi orbitals,\cite{Luken1982_265} \revision{obtained from for spin-$\sigma$ position eigenstates $|\textbf{a}_i^\sigma\rangle$ localized at the so-called Fermi-orbital descriptor (FOD) $\textbf{a}_i^\sigma$ which become the optimized variables in the theory.
When the occupied orbitals are known as a basis set expansion
\begin{equation}
    \psi^\sigma_j (\textbf{r}) = \sum_{\mu}^{M} C_{\mu i}^{\sigma} \chi_{\mu}(\textbf{r}),
    \label{eq:lcao}
\end{equation}
where $\boldsymbol{C}^\sigma$ are the molecular orbital coefficients and $\chi_\mu$ are the basis functions, the corresponding Fermi orbital (FO) coefficients $\boldsymbol{c}_\text{FO}^\sigma$ are given by
\begin{equation}
\boldsymbol{c}_\text{FO}^\sigma = \boldsymbol{C}^\sigma \boldsymbol{R}^{\,\sigma}, 
\label{eq:foloc2}
\end{equation}
where the transformation matrix $\boldsymbol{R}^{\,\sigma}$ is given by
\begin{equation}
	R^{\,\sigma}_{ji} = \frac{\braket{\psi^\sigma_j|\textbf{a}^\sigma_i}}{\sqrt{n^\sigma ({\bf a}_i^\sigma)}} \; . \label{eq:rotation}
\end{equation}
The FOs are non-orthonormal but can be symmetrically orthogonalized\cite{Lowdin1950_365} to yield the Fermi--L\"{o}wdin orbitals (FLOs) with coefficients
\begin{equation}
	\boldsymbol{c}^\sigma = \boldsymbol{c}_\text{FO}^\sigma [\boldsymbol{T}^\sigma  (\boldsymbol{Q}^\sigma)^{-1/2} (\boldsymbol{T}^{\sigma})^{\text{T}} ],
\label{eq:unitarytrafo}
\end{equation}
where $\boldsymbol{T}^{\sigma}$ and $\boldsymbol{Q}^\sigma$ contain the eigenvectors and eigenvalues of the FO overlap matrix, respectively.
These FLOs are then used to evaluate the PZ functional, \eqref{pz}.
} In the spin-restricted formalism, the Fermi orbitals are parametrized by a set of $3N_\text{occ}$ Fermi orbital descriptors (FODs), which correspond to semiclassical cartesian coordinates of the $N_\text{occ}$ electrons of each spin that need to be optimized instead of the $\mathcal{O}(N_\text{occ}^2)$ rotation angles of the traditional formalism.\cite{Pederson1984_1972}
Analytical energy gradients for the FODs have been derived in \citeref{Pederson2015_064112}, and they are employed in various implementations of the FLO-SIC method\revision{; please refer to \citet{Schwalbe2020_084104} for a thorough overview on the presently used, openly available implementation of FLO-SIC}.

\citet{Schwalbe2019_2843} showed that the FODs and the corresponding FLOs have a clear relation to chemical bonding theories.
The most common bond patterns in LT are single, double, and triple bonds, and any of these bonds can be represented in terms of FOD arrangements.
For instance, single bonds are described by one $\alpha$ and one $\beta$ FOD in-between the atoms, double bonds are characterized by two $\alpha$ and two $\beta$ FODs 
above and below the bond center, and triple bonds are built up by three $\alpha$ and three $\beta$ FODs placed in triangles around the bond center for each spin.
Analogous FOD motifs have also been found for FODs corresponding to core and lone pair orbitals:\cite{Schwalbe2019_2843} for example, FODs corresponding to 1s orbitals are typically placed at the nuclear sites, while the FODs for fully occupied 2s and 2p orbitals usually form a tetrahedron.

\begin{figure}[h]
    \centering
    \includegraphics[width=\linewidth]{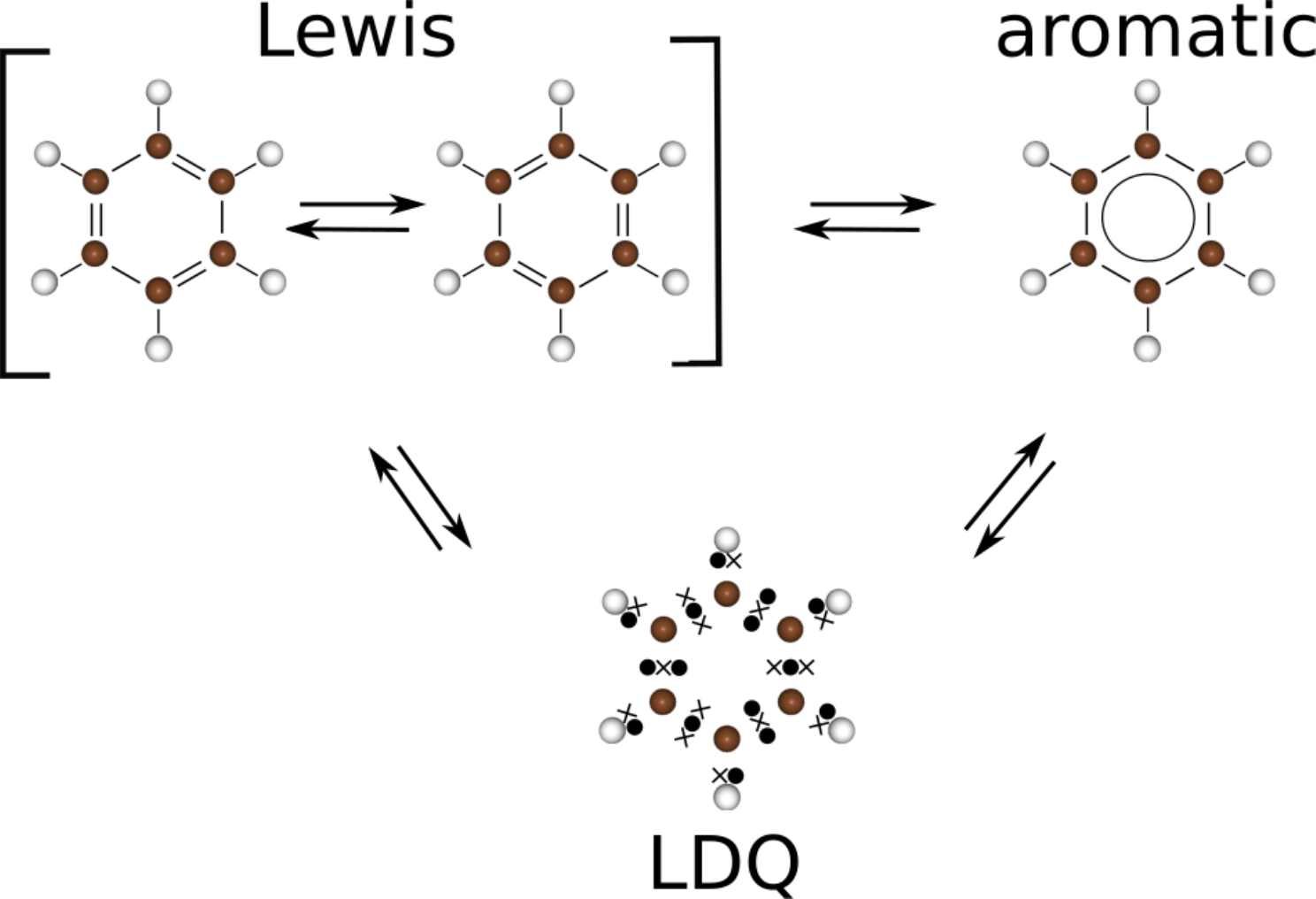}
     \caption{Electronic geometries of benzene according to various bonding theories. A \emph{single} LT structure cannot describe the aromaticity correctly; only a superposition of the two resonance structures reproduces the aromatic character. On the other hand, a \emph{single} LDQ structure already describes the electronic configuration correctly\revision{, since the spin-average of the electronic geometry has a formal bond order of 1.5 in all bonds}.
     Solid lines represent a bond involving one spin-up and one spin-down electron. In LDQ theory, the x symbol denotes a spin-up electron while \revision{the $\bullet$ symbol} denotes a spin-down electron. The arrows indicate transformations between the representations of aromatic bonding. 
       }
    \label{fig:benzene_intro}
\end{figure}

While LDQ theory includes all the bonding motifs of LT, it also allows a simple and elegant representation of aromatic bonds.
As illustrated by \figref{benzene_intro}, an aromatic bond can be represented in LDQ by two $\alpha$ FODs and one $\beta$ FOD or vice versa, leading to a formal bond order of 1.5, whereas the LT structures have alternating single and double bonds.

In this work, we investigate these findings further by studying the usefulness of FODs that follow the bonding theories of Lewis and Linnett for exploring symmetry breaking in SIC calculations.
Symmetry breaking in SIC was originally discovered by \citet{Lehtola2016_3195}, who showed that the molecular symmetry of benzene is broken by SIC with real (RSIC) or complex orbitals (CSIC).
As discussed in \citeref{Lehtola2016_3195}, the symmetry breaking arises because SIC calculations on benzene are based on a single LT structure with alternating single and double bonds.
The energy of the SIC solution can be decreased by allowing the C-C double bonds in the Lewis structure to contract, leading to more spatial localization which is energetically favored by SIC, while the C-C single bonds of the Lewis structure become elongated.
The carbon-carbon bonds in benzene then come out unequal in length, similarly to (but not as strongly as in) the fictitious 1,3,5-cyclohexatriene molecule.

While SIC solutions guided by LT thus break the molecular symmetry of benzene, we show in this work that LDQ theory succeeds in reproducing a symmetric molecular geometry and thereby overcomes symmetry breaking for this molecule.
We also show that the symmetry breaking of LT can already be seen in the electron density at the symmetric molecular geometry of benzene.
To highlight the broader scope of these findings for various kinds of SIC, in this work we consider not only the FLO-SIC method, but also include variational SIC in both the RSIC and CSIC variants to show how the choice of the initial orbital guess affects self-consistent SIC solutions.

Moreover, as artifacts arising from the use of a single Lewis structure may also become visible in the electron density, we point out the need to carefully analyze SIC solutions. 
Following the pioneering work of \citet{Hait2018_1969}, who suggested using electronic dipole moments (DMs) as a simple global measure of the electron density reproduced by DFT, we investigate with DMs whether the electron density from SIC calculations properly reflects the molecular symmetry.
Using a benchmark set of 16 planar cyclic molecules, which extends a database previously suggested by \citet{Adhikari2020_184303}, we show that bonding patterns guided by LT lead in some cases to SIC DMs that do not reflect the molecular symmetry, leading to artifactual dipole moments.
For example, in the case of phenazine, \ce{(C6H4)2N2}, which is symmetric and therefore should not have a dipole moment, LT structures result in a huge artifactual dipole moment of 1~Debye.

Motivated by the success with benzene, we also show how LDQ can be used to find alternative bonding patterns that lead to improved DMs in SIC calculations for this database.
The LDQ structures for all the studied molecules have the same symmetries as the molecules themselves, at variance to the LT structures.
The correct symmetry of the FODs also leads to DMs with the correct symmetry; in the previously discussed case of phenazine, LDQ correctly yields no dipole moment.
Based on this victory, we point out that LDQ FODs can be seen as a simple way to model aromatic molecules in SIC. 
We elaborate that FLOs based on LDQ FODs, which can be interpreted as a superposition of LT structures, can be used as initial guesses for RSIC and CSIC, leading to stable local minima.

This work is structured as follows. 
The methodology used in this study is discussed in \secref{methods}, and details of the electronic structure calculations carried out in this work are provided in \secref{computational}.
The results are presented in \secref{results}.
First, various local minima and symmetry breaking in benzene are discussed in detail using FLO-SIC, RSIC, and CSIC in \secref{benzene_results}. 
Next, the influence of the initial orbital guesses following LT or LDQ arrangements on the DMs arising from SIC calculations on our benchmark set of planar, cyclic molecules is discussed in \secref{dip16_results}, where calculated DMs are used to explain how the SIC solutions are affected by the chemical bonding situations described by the initial orbital guess; basis set convergence is also investigated. Further investigations into bonding motifs are discussed in \secref{further}.
The main findings of this study are summarized and conclusions are drawn in \secref{summary}.

\section{Methods}
\label{sec:methods}

\subsection{Benzene study}
\label{sec:benzenemethod}

\begin{figure}
 \includegraphics[width=0.45\textwidth]{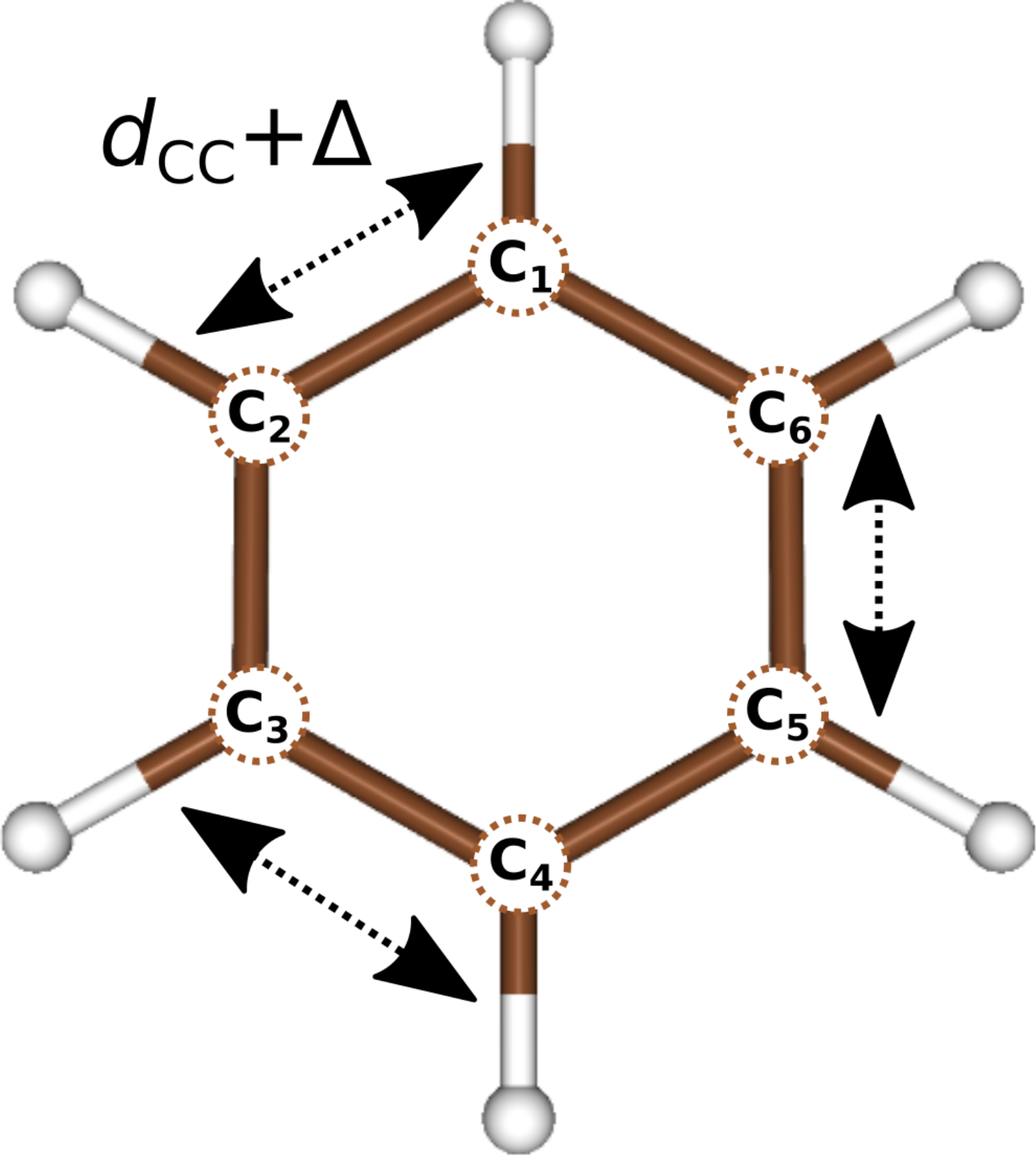}
 \caption{Nuclear geometry of benzene, showing the bond lengths
   $d_\text{CC}$ which are elongated by the addition of the local distortion
   $\Delta$. The other three bonds are shortened accordingly.}
 \label{fig:nuclear_benzene_model}
\end{figure}

Our goal in this section is to study the effect of various electronic geometries, that is, FOD arrangements, on the molecular structure of benzene.
Benzene is the prototypical aromatic molecule, and as will be shown later in this work, any conclusions that can be derived for benzene are likely applicable also to larger aromatic systems as well.
As has been already established,\cite{Lehtola2016_3195} SIC prefers a symmetry-broken structure for benzene; the energy landscape of benzene was therefore studied as follows.
Average \mbox{C-C} bond lengths $d_{\text{CC}}$ in the range between 1.317~\AA{} and 1.417~\AA{} were sampled with a step size of 0.005~\AA{}, yielding 21 values for the average bond length; experimentally, all the C-C bond 
lengths are the same, $d_{\text{CC}}=1.397$~\AA{}.\cite{CCCBDB} 
All C-H bonds were fixed to their experimental value, $d_{\text{CH}} = 1.084$~\AA{}.\cite{CCCBDB}
To investigate the distorted structures that may be favored by SIC, following \citeref{Lehtola2016_3195} we elongate three \mbox{C-C} bonds by an amount $\Delta$, whereas the other three \mbox{C-C} bonds are shortened; the hydrogens are moved correspondingly.
We consider local distortions $\Delta$ ranging from 0~\AA{} (symmetric structure) to $0.075$~\AA{}, likewise sampled in 0.005~\AA{} steps, leading to 16 distorted structures for each value of the average C-C bond length and a grand total of 336 candidate structures.
A detailed description of how to obtain the nuclear coordinates with this procedure is given in the supporting information (SI); for a visual explanation, see \figref{nuclear_benzene_model}.

\begin{figure}
\centering
\hfill \subfigure[LT1]{\includegraphics[width=0.155\textwidth]{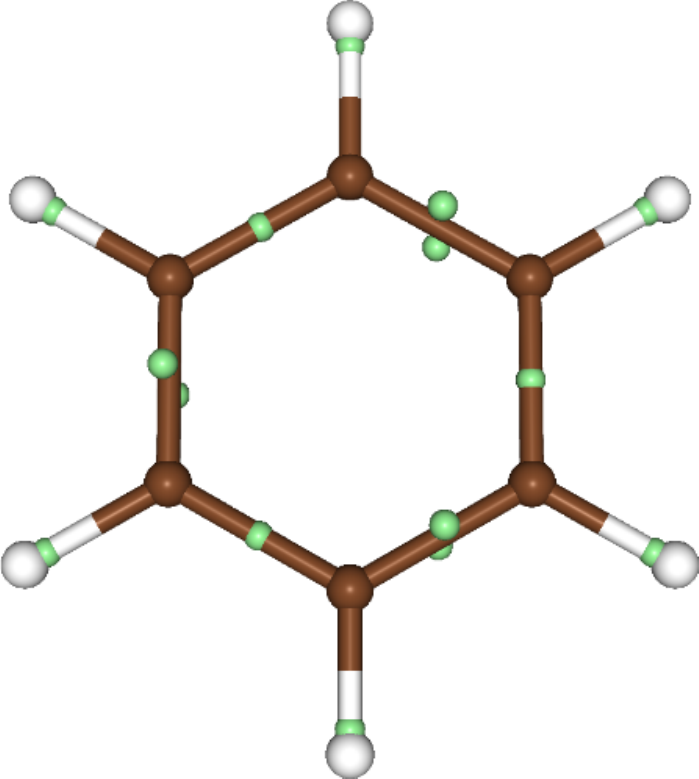}} \hfill \subfigure[LT2]{\includegraphics[width=0.155\textwidth]{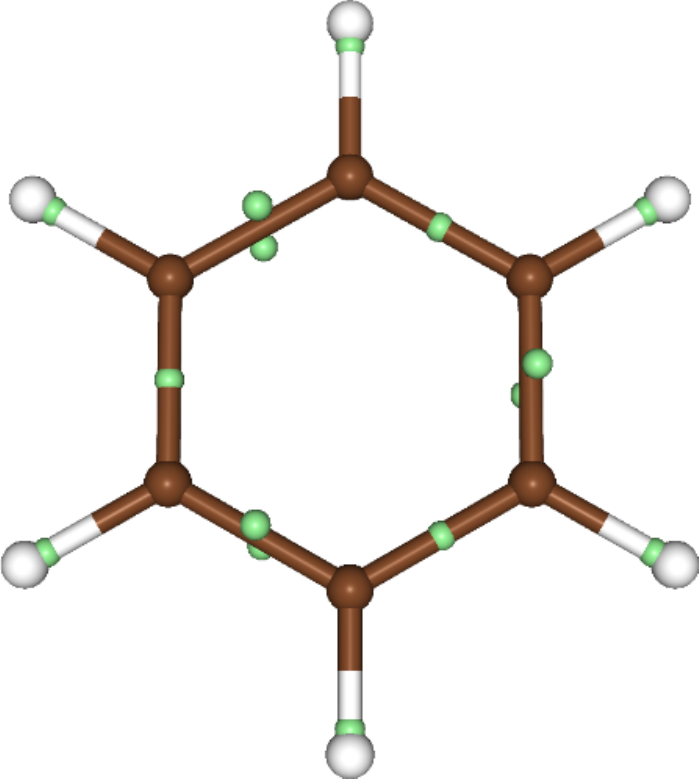}} \hfill \subfigure[LDQ]{\includegraphics[width=0.155\textwidth]{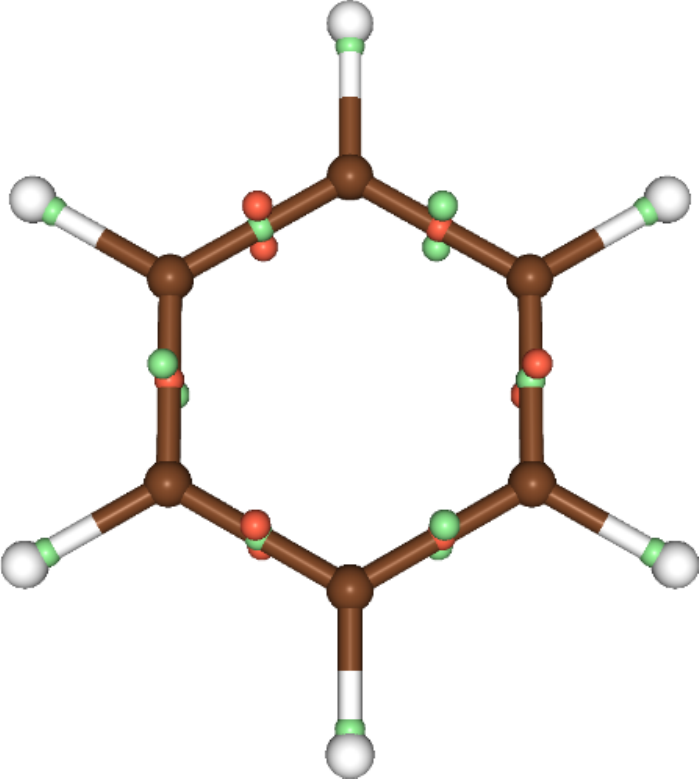}} \hfill
 \caption{The FOD arrangements for benzene considered in this study.
 Carbons are shown in brown, hydrogens in white, spin-up FODs in red, and spin-down FODs in light green.
 The spin-up and spin-down FODs are identical in cases where the spin-up FODs are not visible.
 The LT1 and LT2 structures differ in the placement of the single and double bonds, while the LDQ structure takes the spin-up FODs from LT2 and the spin-down FODs from LT1.
 (There is also another LDQ structure in which the roles of LT1 and LT2 are interchanged; however, it was not considered in this study as it leads to identical results.)} 
 \label{fig:figure_benzene_configs}
\end{figure}

The three FOD starting structures (LT1, LT2, and LDQ) shown in \figref{figure_benzene_configs} were considered for benzene.
In the LT1 structure, the double-bond FODs are placed between C$_{2}$-C$_{3}$, C$_{4}$-C$_{5}$ and C$_{1}$-C$_{6}$; the numbering is shown in \figref{nuclear_benzene_model}. 
In the LT2 structure, the double bonds are placed between \mbox{C$_{1}$-C$_{2}$}, C$_{3}$-C$_{4}$ and C$_{5}$-C$_{6}$, instead.
The LDQ structure is staggered: a double-bond FOD structure is picked for one spin channel, while a single-bond FOD structure is used for the other spin channel, and the configurations alternate spins between each bond.
The LDQ structure thereby formally corresponds to a broken-symmetry open-shell singlet state; however, it correctly yields a formal bond order of 1.5 for all C-C bonds which are expected to be aromatic.

\subsection{DIP16 benchmark set}
\label{sec:dip16}

The benchmark set used within this study extends an existing benchmark set introduced by \citet{Adhikari2020_184303}; a full list of the molecules is given in \tabref{DIP16} along with the corresponding shorthand identifiers used in this work.
The original benchmark set of \citeref{Adhikari2020_184303} consists of 14 molecules containing 44 to 120 electrons and H, C, N, O, F, S, and Cl atoms.
All systems in the set are cyclic and can be divided into aromatic and non-aromatic molecules.
\citet{Adhikari2020_184303} only compared the DMs of 10 of their 14 molecules to experimental values (see \tabref{DIP16}). 
In this work, we have included the experimental DMs for the four missing systems: 2,1,3-benzothiadiazole, dichlone, fluorene, and 1,2,5-thiadiazole.
In addition, we have included cyclobutadiene to account for anti-aromaticity as well as benzene for a fuller coverage of various chemistries.
We will refer to this entire set as DIP16 (DIPoles of 16 molecules).

In the interest for maximal comparability to the study of \citet{Adhikari2020_184303}, the molecular structures used in this work were collected from common chemical databases \cite{NISTwebbook,chemspider,Kim2020_D1388};
an explicit listing is given in the SI. 
All the used molecular and FOD structures as well as the corresponding DMs are also available at \url{https://gitlab.com/kaitrepte/dip16}. 

\begin{table} 
    \centering
    \caption{Description of the DIP16 benchmark set, including the molecules' identifiers (IDs) and point groups (PGs). More details can be found in the SI. \\ $^{a}$ Molecules studied by \citet{Adhikari2020_184303} but whose DMs were not compared to experiment in  \citeref{Adhikari2020_184303}. \\
    $^{b}$ Molecules introduced in this work.}
    \footnotesize
    \begin{tabular}{l|llll} %
         ID             & PG     & Molecule               & Formula                       & Group          \ \\ \hline \hline
         mol01          & \CTv{} & acridine               & C$_{13}$H$_{9}$N              & aromatic       \\
         mol02          & \DTh{} & anthracene             & C$_{14}$H$_{10}$              & aromatic       \\
         mol03          & \CTv{} & azulene                & C$_{10}$H$_{8}$               & aromatic       \\
         mol04          & \DTh{} & p-benzoquinone         & C$_{6}$H$_{4}$O$_{2}$         & non-aromatic   \\
         mol05$^{a}$    & \CTv{} & 2,1,3-benzothiadiazole & C$_{6}$H$_{4}$N$_{2}$S        & aromatic       \\ 
         mol06          & \Cs{}  & benzothiazole          & C$_{7}$H$_{5}$NS              & aromatic       \\  
         mol07          & \DTh{} & p-chloranil            & C$_{6}$Cl$_{4}$O$_{2}$        & non-aromatic   \\  
         mol08$^{a}$    & \CTv{} & dichlone               & C$_{10}$H$_{4}$Cl$_{2}$O$_{2}$ & non-aromatic  \\ 
         mol09          & \DTh{} & p-fluoranil            & C$_{6}$F$_{4}$O$_{2}$         & non-aromatic   \\ 
         mol10$^{a}$    & \CTv{} & fluorene               & C$_{13}$H$_{10}$              & non-aromatic   \\ 
         mol11          & \CTv{} & 1,4-naphthalenedione   & C$_{10}$H$_{6}$O$_{2}$        & non-aromatic   \\ 
         mol12          & \DTh{} & phenazine              & C$_{12}$H$_{8}$N$_{2}$        & aromatic       \\
         mol13$^{a}$    & \CTv{} & 1,2,5-thiadiazole      & C$_{2}$H$_{2}$N$_{2}$S        & aromatic       \\ 
         mol14          & \CTv{} & thiophene              & C$_{4}$H$_{4}$S               & aromatic       \\  
         mol15$^{b}$    & \DTh{} & cyclobutadiene         & C$_{4}$H$_{4}$                & anti-aromatic  \\
         mol16$^{b}$    & \DSh{} & benzene                & C$_{6}$H$_{6}$                & aromatic       \\
    \end{tabular}
    \label{tab:DIP16}
\end{table}

For the entire DIP16 test set, we investigated a total of 51 LT and LDQ electronic geometries as initial starting points for RSIC, CSIC, and FLO-SIC calculations.
There are up to 4 LT and 2 LDQ electronic geometries for each molecule; an example will be provided in \secref{results}, 
and some more visualizations are available in the SI.
The various LT and LDQ configurations for each molecule lead to distinct DMs, as the total density that arises from the orbital optimization is biased by the bonding pattern described by the FOD configuration.
The DM components of DFT, RSIC, CSIC, and FLO-SIC calculations are also detailed in the SI.

\section{Computational details}
\label{sec:computational}
\subsection{Electronic structure codes}
The Gaussian-basis all-electron codes
\ERKALE{}~\cite{Lehtola2012_1572, Hel2016_1, Lehtola2014_5324, Lehtola2016_3195},
\PySCF{}~\cite{Sun2020_024109}, and \PyFLOSIC{}~\cite{Schwalbe2020_084104}
were used in this work. 
As \ERKALE{} and \PyFLOSIC{} are based on similar computational approaches (both employ the standard approach used in almost all Gaussian-basis quantum chemistry programs, which allows routinely reproducing total energies to microhartree accuracy), they enable comparisons of FLO-SIC, RSIC, and 
CSIC on an equal footing.
The codes used in this work have a focus on simple input and output file handling as well as user-friendliness, which makes setting up complex workflows rather easy.
Moreover, all three codes are freely and openly available, and can be used without limitations also in industry; see \citeref{Lehtola2021__} for thorough discussion.

Furthermore, all three codes evaluate density functionals with the open-source \textsc{Libxc} library,\cite{Lehtola2018_1} again ensuring comparability of the results. 
The Perdew--Wang (PW92) correlation functional,\cite{Perdew1992_13244} which is a local density approximation (LDA),  has been used for all calculations described in this work in combination with LDA exchange,\cite{Bloch1929_545, Dirac1930_376} if not stated otherwise.

Even though the FODs depend on the exchange-correlation functional, in analogy to the findings of \citeref{Lehtola2016_3195}, the results obtained at the LDA level of theory with the PW92 functional suffice to demonstrate symmetry-breaking effects arising from the self-interaction correction.
Indeed, as will be demonstrated in \secref{methodvalidation}, calculations performed at the generalized gradient approximation (GGA) as well as the meta-GGA levels of theory yield results that are similar to those obtained at the LDA level with the PW92 functional.

The all-electron calculations in \ERKALE{}, \PySCF{} and \PyFLOSIC{} were carried out with the double-$\zeta$ polarization consistent pc-1 basis set.\cite{Jensen2001_9113, Jensen2002_7372}
While DMs are a textbook example of the need for diffuse basis functions,\cite{Lehtola2012_104105} as the expectation value of the dipole moment operator $\langle \hat{\bf r} \rangle$ may accumulate significant contributions from small changes in the electron density at large ${\bf r}$ where only diffuse functions contribute, the pc-1 basis set was chosen for comparability to the work of \citet{Adhikari2020_184303} that employed the DFO basis set\cite{Porezag1999_2840} which is limited to $d$ functions, thereby contains only a single shell of polarization functions on main group elements, and is therefore formally of polarized double-$\zeta$ quality; this is also supported by numerical evidence.\cite{Schwalbe2020_084104}
Although the use of such small basis sets is generally not advisable for investigating properties as challenging as DMs, exploratory calculations carried out with larger basis sets in \secref{methodvalidation} show that the findings of the present study are not compromised by the use of such a small basis set that was chosen to optimize the computational cost.

\subsubsection{PySCF and PyFLOSIC}
\label{sec:pyflosic_comp}
A grid level\cite{Treutler1995_346} of 7 was used in \PySCF{} and \PyFLOSIC{}. This corresponds to using a (90,974) grid for hydrogen, a (135,1202) grid for second-period atoms and a (140,1202) grid for third-period atoms in Becke's multicenter quadrature scheme.\cite{Becke1988_2547}
In contrast to the total electron density in normal DFT calculations, the orbital densities considered in SIC are not smooth; this is the main reason for the need for huge integration grids in SIC calculations, and also the rationale for turning off grid pruning in all DFT and SIC calculations of this work.
Any linear dependencies in the molecular atomic-orbital basis set were removed by applying the \texttt{scf.addons.remove\_linear\_dep\_} function on the \PySCF{} DFT object; default \PySCF{} thresholds for this procedure were used.
An SCF energy criterion of $10^{-8}~E_{\text{h}}$ was used in all calculations.

Starting values for the FODs for \PyFLOSIC{} calculations were generated with the Fermi-orbital descriptor Monte-Carlo \fodMC{}~code\cite{Schwalbe2019_2843} (\url{https://github.com/pyflosic/fodMC}).
While there are other ways to initialize the FODs (see \citerefs{Schwalbe2019_2843} and \citenum{Schwalbe2020_084104} for discussion), the \fodMC{} code enables a direct translation between specific chemical bonding situations and FOD guesses, making \fodMC{} the ideal approach to study the differences between LT and LDQ within SIC.
The starting FODs were optimized in the FLO-SIC calculations performed with \PyFLOSIC{} until the maximum force $\max_i|{\bf F}_i|$ acting on any FOD fell below $5\cdot10^{-4}$~$E_{\text{h}}$/$a_{0}$.

As it is well-known that a vanishing gradient does not necessarily imply convergence onto a local minimum, for verification purposes, we introduce a stability analysis of FLO-SIC solutions in analogy to the method introduced in \citeref{Lehtola2016_3195} for RSIC and CSIC calculations.
A FLO-SIC calculation will be deemed to have converged onto a stable solution if the FOD Hessian is positive semidefinite.
In this work, the FOD Hessian is formed seminumerically via finite differences of analytical FOD forces, using a two-point stencil with a step size of 0.001~\AA{}.
Because of the small number of FODs, the FOD Hessian fits in memory even for large systems and can be diagonalized using dense matrix algebra.

\subsubsection{ERKALE}
\label{sec:erkale_comp}
An unpruned (150,1202) quadrature grid was used in \ERKALE{}.
The threshold for the gradients of the orbital rotations in the occupied-occupied block was set to $10^{-5}~E_{\text{h}}$.

Stability analyses\cite{Lehtola2016_3195}---checks that the SIC solutions correspond to local minima with respect to rotations of the occupied orbitals---were carried out for all RSIC and CSIC calculations; the CSIC calculations were started from stable RSIC minima according to the best practices established in \citeref{Lehtola2016_3195}.

The default initial orbital guess in \ERKALE{} for RSIC and CSIC calculations is given by localized orbitals\cite{Lehtola2013_5365} according to the FB criterion.
To allow starting RSIC calculations from the same initial guess as \PyFLOSIC{}, \ERKALE{} was extended in this work to be able to initialize RSIC and CSIC calculations from FODs.
Thus, instead of performing a FB localization to get initial orbitals for SIC calculations, input FODs are read in alongside a converged DFT wave function; FLOs are then constructed from these data \revision{using Eqs.~(\ref{eq:foloc2})--(\ref{eq:unitarytrafo})}.
\revision{After the initialization, the Perdew--Zunger functional, \eqref{pz}, is minimized as usual by variational optimization, using the orbital rotation technique detailed in \citeref{Lehtola2016_3195}}.

\section{Results}
\label{sec:results}

\subsection{Benzene study}
\label{sec:benzene_results}

As was discussed in \secref{benzenemethod}, benzene has three distinct FOD geometries: LT1, LT2, and LDQ, which were already shown in \figref{figure_benzene_configs}.
Because LT1 and LT2 only differ by a rotation of the molecule, it is expected that both LT configurations converge onto equivalent minima.\cite{Lehtola2016_3195}
The LT structures are expected to yield symmetry-broken geometries, such that the double-bond FODs prefer shorter C-C bonds. 
On the other hand, as the LDQ configuration can be thought of as a superposition of the two LT configurations, in which the spin-up FODs are picked from one LT structure while the spin-down FODs are picked from the other LT structure, the LDQ structure is expected to prefer a symmetric molecular geometry.

\begin{table}[h]
\centering
\caption{Energetically preferred solutions from different methods, using \ERKALE{} and \PySCF{}/\PyFLOSIC{} for benzene.
The shorter C-C bond length, $d_{\text{CC}}^{\text{short}}$, the longer C-C bond length $d_{\text{CC}}^{\text{long}}$, the total energy $E_{\text{tot}}$ and the $\braket{\hat{\bf S}^{2}}$ value are shown. 
If the two types of bonds have the same length, the calculation prefers a symmetric geometry.
The PW92 exchange-correlation functional was used in all calculations.
  }
\begin{tabular}{l|rrrrr}
Code                        & Method    & $d_{\text{CC}}^{\text{short}}$ [\AA{}]  & $d_{\text{CC}}^{\text{long}}$ [\AA{}]    & $E_{\text{tot}}$ [$E_{\text{h}}$] & $\braket{\hat{\bf S}^{2}}$  \\\hline
\multirow{1}{*}{\PySCF{}}   & DFT       & 1.392                                 & 1.392            & $-230.042959$     & 0.000        \\\hline
\multirow{4}{*}{\PyFLOSIC{}}& \multicolumn{5}{c}{FLO-SIC} \\
                            & LT1       & 1.322                                 & 1.405            & $-233.023405$     & 0.000       \\ 
                            & LT2       & 1.322                                 & 1.405            & $-233.023405$     & 0.000       \\ 
                            & LDQ       & 1.362                                 & 1.362            & $-233.017015$     & 0.150       \\\hline
                            
\multirow{11}{*}{\ERKALE{}} & DFT       & 1.392                                 & 1.392            & $-230.042959$     & 0.000        \\
                            
                            & \multicolumn{5}{c}{RSIC} \\
                            & FB        & 1.322                                 & 1.405            & $-233.031182$     & 0.000        \\         
                            & LT1       & 1.322                                 & 1.405            & $-233.031182$     & 0.000        \\
                            & LT2       & 1.322                                 & 1.405            & $-233.031182$     & 0.000        \\
                            & LDQ       & 1.362                                 & 1.362            & $-233.024906$     & 0.148        \\  
                    
                            & \multicolumn{5}{c}{CSIC} \\
                            & FB        & 1.337                                 & 1.390            & $-233.082400$     & 0.000        \\
                            & LT1       & 1.337                                 & 1.390            & $-233.082394$     & 0.000        \\
                            & LT2       & 1.337                                 & 1.390            & $-233.082394$     & 0.000        \\ 
                            & LDQ       & 1.362                                 & 1.362            & $-233.079975$     & 0.098        \\\hline
                            
\end{tabular}
\label{tab:benzene_results}
\end{table}

The bond lengths and total energies for DFT, RSIC, CSIC, and FLO-SIC are shown in \tabref{benzene_results}. 
While DFT predicts a symmetric molecule, RSIC and CSIC break the symmetry as found by \citet{Lehtola2016_3195}. 
For RSIC, we find C-C bond lengths of 1.322~\AA{} and 1.405~\AA{}, while for CSIC the bond lengths are 1.337~\AA{} and 1.390~\AA{}, respectively.
As expected, these results are in excellent agreement with the ones of \citeref{Lehtola2016_3195} that found bond lengths of 1.32~\AA{} and 1.40~\AA{} for RSIC and 1.33~\AA{} and 1.39~\AA{} for CSIC with the cc-pVTZ basis set and the PW92 functional, respectively.
This result serves as a consistency check for our further discussion. Furthermore, the DFT energies produced by \PySCF{} and \ERKALE{} are identical, showing that the calculations performed with the two codes can be meaningfully compared.

Moving onto FLO-SIC, the calculations with \PyFLOSIC{} started from LT type guesses give analogous results to the RSIC calculations with \ERKALE{}: the molecular structure is distorted, as the minimal-energy structure corresponds to short double bonds and long single bonds.
Although FLO-SIC expectedly produces a higher total energy than RSIC, as the latter method has more optimizable degrees of freedom as it allows all possible real-valued occupied-occupied orbital rotations, the optimal bond lengths from RSIC and FLO-SIC match to the precision used in this study.

In contrast, the electronic geometry from LDQ leads to a symmetric molecular structure. 
The solution obtained from the LDQ initial guess is slightly higher in energy (by about 6~m$E_{\text{h}}$) than that from the LT initial guesses in the RSIC calculations in \ERKALE{}; a similar energy difference is also observed in the FLO-SIC calculations in \PyFLOSIC{}.  
The energy difference of the LT and LDQ solutions is smaller in the CSIC calculations performed with \ERKALE{}, measuring about 2.5~m$E_{\text{h}}$.
However, these millihartree differences are negligible compared to the overall magnitude of the SIC, which is in the order of 3~$E_\text{h}$ as evidenced by the data in \tabref{benzene_results}.

The C-C bond length optimized with DFT and the PW92 functional is in excellent agreement with the experimental bond length 1.397~\AA{}, with a difference of only 0.005~\AA{}, which is also the resolution of this study.
The LDQ calculation underestimates the experimental bond length by about 0.035~\AA{}, whereas the LT structures break the symmetry, predicting one bond to be 0.075~\AA{} shorter and the other to be 0.008~\AA{} longer than in the experiment.
As these results show, not only does LDQ resolve the symmetry breaking, but the maximal error in the bond lengths is two times smaller than in the LT structures.
Interestingly, the optimal bond length for the LDQ structure, 1.362~\AA{}, is close to the average of the optimal bond lengths for the LT structures, 1.364~\AA{}.

As can be expected from the correspondence of the LDQ structure to broken-symmetry open-shell singlet calculations, the spin symmetry is slightly broken as evidenced by the non-zero $\braket{\hat{\bf S}^{2}}$ from the LDQ calculations using RSIC, CSIC and FLO-SIC.
In contrast, there is no spin contamination in the LT structures, but they do break the molecular symmetry.

\begin{figure}
 \includegraphics[width=0.45\textwidth]{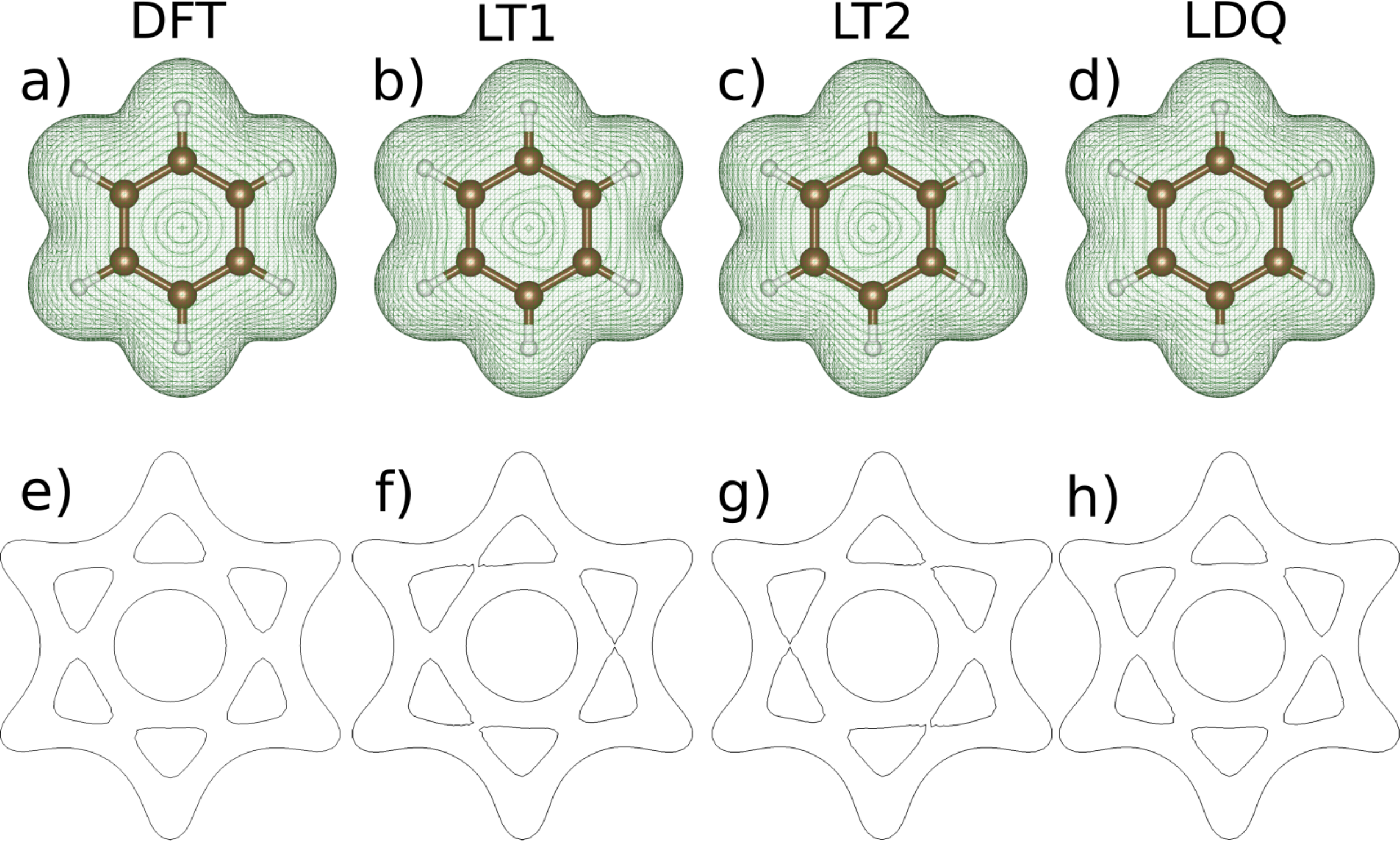}
 \caption{Total electron density in benzene according to DFT as well as RSIC initialized with FLOs following an LT or LDQ structure.
 Subfigures (a--d) show a wireframe representation of the densities with an isosurface value of 0.004 electrons/$a_0^3$.
 Subfigures (e--h) show the corresponding two-dimensional contour plots of the density at $z=0.4$~\AA{} above the molecular plane.
 }
 \label{fig:density_benzene}
\end{figure}

\begin{figure}
 \includegraphics[width=0.45\textwidth]{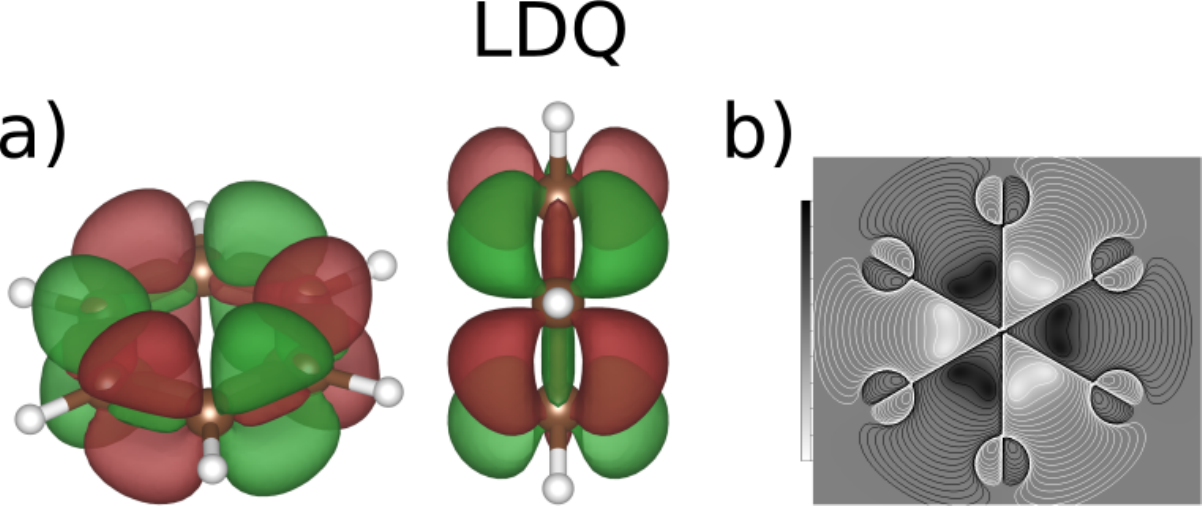}
 \caption{Spin density of benzene from the LDQ solution. a) Isosurfaces of the spin density from two perspectives. b) Two-dimensional contour plots of the spin density at $z=0.4$~\AA{} above the molecular plane.}
 \label{fig:spin_density_benzene}
\end{figure}

The electron densities resulting from DFT and the RSIC solutions guided by the LT1, LT2 and LDQ FOD starting guesses in \ERKALE{} are shown in \figref{density_benzene}.
As expected, DFT delivers a symmetric electron density. 
Strikingly, symmetry breaking is already observable in the electron density of the two LT solutions at the symmetric molecular structure reproduced by DFT (see \tabref{benzene_results}):
both LT1 and LT2 break the rotational symmetry by placing more electron density onto the double bonds of the corresponding Lewis structure.

LDQ restores a rotationally invariant electron density, like the one from DFT.
However, as expected from the open-shell singlet character of LDQ evidenced by the the non-zero value of $\langle \hat{\bf S}^2 \rangle$, the LDQ calculation results in a spin density, which is shown in \figref{spin_density_benzene}: the spin-alternation of the single and double bond FODs in the LDQ structure (see \figref{figure_benzene_configs}) shows up as an alternating spin density.

\subsection{DIP16}
\label{sec:dip16_results}

Having exemplified the existence of several possible types of SIC solutions for benzene, we will next examine the DMs arising from various types of SIC solutions for our benchmark database of 16 planar cyclic molecules.
In the interest of keeping the discussion short, the large set of data is only discussed summarily in the main text; thorough analyses of all molecules can be found in the SI.
Note again that as discussed in \secref{methods}, all calculations in this section are performed at fixed geometries, at variance to the calculations on benzene discussed in \secref{benzene_results}.

\subsubsection{Aromatic molecules}
\label{sec:aromatic_DIP16}

\begin{figure}[h]
    \centering
    \includegraphics[width=\linewidth]{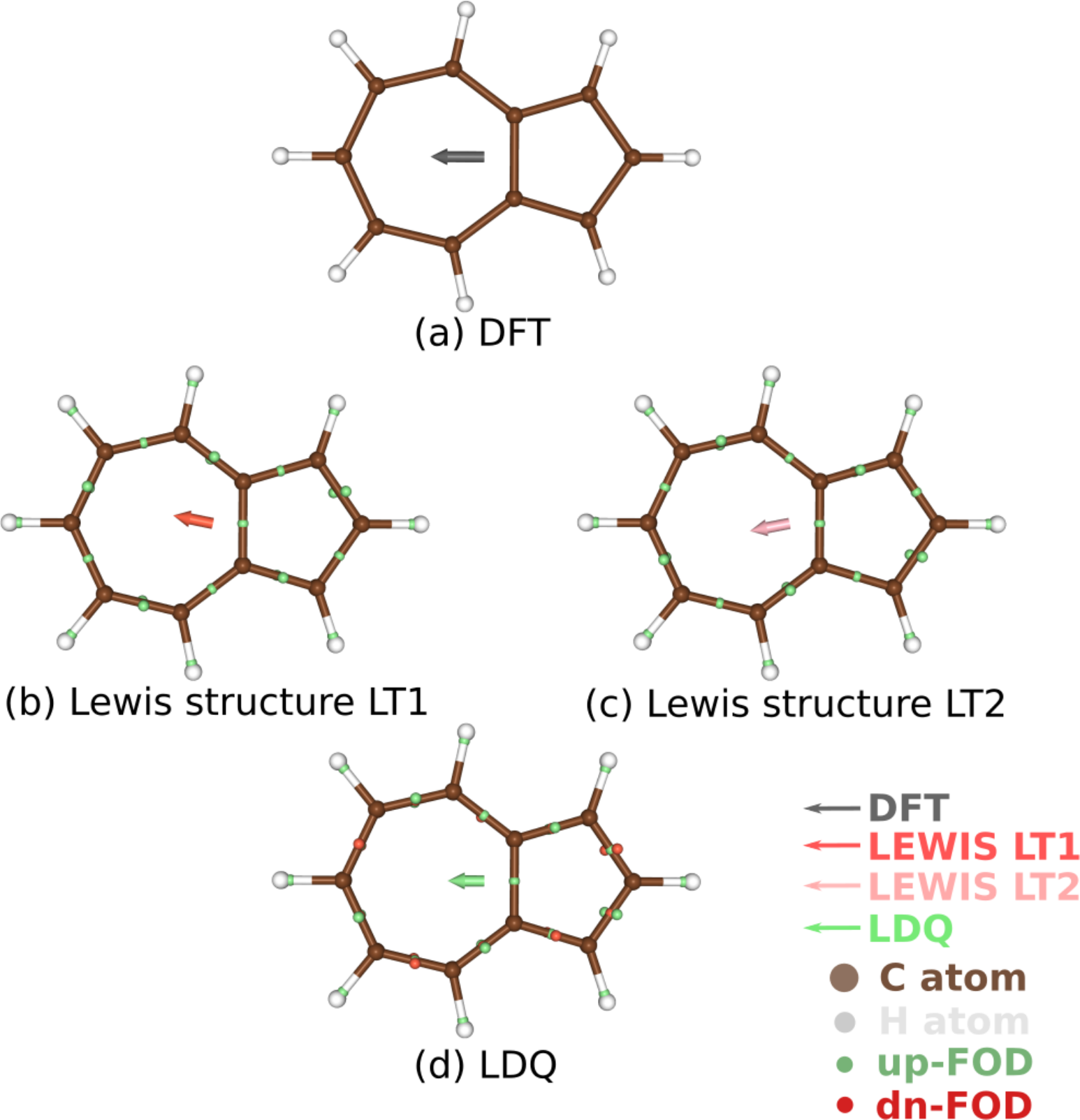}
     \caption{Azulene (mol03): DMs of stable RSIC solutions from \ERKALE{}, starting
       from various initial FOD geometries. The resulting DMs are
       clearly different: only the ones from DFT and SIC
         guided by LDQ are in agreement with the molecular symmetry.
         DMs from RSIC and FLO-SIC point in the
       same direction, and their $\mu$ values agree well for each FOD configuration.}
    \label{fig:erkale_mol03_dips}
\end{figure} 

\figref{erkale_mol03_dips} shows the DMs of azulene (mol03, \ce{C10H8}, \CTv{} point group) corresponding to DFT and stable RSIC solutions from \ERKALE{}. 
The calculations for \figref{erkale_mol03_dips} were started from FLOs generated by three distinct sets of FODs, analogously to the ones for benzene discussed in \secref{benzene_results}.
The three FOD starting points (LT1, LT2, and LDQ) each lead to a stable solution in \ERKALE{}, as verified by the stability analysis with respect to rotations of the occupied orbitals discussed in \secref{erkale_comp}.
Stability analysis was also employed in the FLO-SIC calculations on this molecule (see \secref{pyflosic_comp}), and the three solutions were each found to be true local minima.

As can be seen from \figref{erkale_mol03_dips}, the DM from SIC started from an LDQ configuration points along the two-fold symmetry axis; this means that the LDQ structure leads to an electron density that reflects the symmetry of the molecule. 
In contrast, the electron densities obtained with SIC starting from LT structures violate this symmetry, and as a result the DMs do not point in the direction of the two-fold symmetry axis. 
The two LT structures are mirror images of one another, and their DMs have an angle of roughly 10$^{\circ}$ with respect to the symmetry axis.
The asymmetry can already be seen from the LT FODs: counting all C-C bond FODs in the symmetry-equivalent parts of the molecule shows that there are 14 FODs on one side, and 16 on the other.
This imbalance is then reflected and reinforced when the electron density is optimized in the self-consistent field procedure, as was already mentioned in \secref{dip16}.

Moreover, as mentioned in \secref{benzenemethod}, combining the spin-up FODs of LT1 and the spin-down FODs of LT2 results in a LDQ structure. 
This LDQ structure has 15 \mbox{C-C} bond FODs in both symmetry-equivalent parts of the molecule, and thereby affords an unbiased description for the molecule.
Thus, we can introduce LDQ structures as a useful representation of aromatic bonding.
The character of the LDQ structure as a superposition of the LT structures is also observable in the DM: the averaged LT DM is $\mu_{\text{(LT1+LT2)/2}} = 0.71$~D, it aligns with the molecular axis like the DMs from DFT and LDQ, and has a similar size to the LDQ DM, $\mu_{\text{LDQ}} = 0.65$~D.

There are several other molecules in the DIP16 set where LT FODs likewise lead to incorrect DMs.
In acridine (mol01), LT configurations can lead to DMs with an angle of about 14$^{\circ}$ with respect to the symmetry axis, while the LDQ DM correctly reflects the symmetry of the molecule.
Phenazine (mol12) has LT structures which introduce a significant DM of about 1~D which is oriented perpendicular to the N-N axis, while the DM should be zero according to the molecule's \DTh{} point group; only the LDQ solution reproduces the correct DM.
Similar results are determined for \PyFLOSIC{}, see the SI, and the observations are analogous. For instance, 
the stability analysis discussed in \secref{pyflosic_comp} was carried out for the FLO-SIC solutions of azulene (mol03), and they were  found to be true local minima.

For either \ERKALE{} or \PyFLOSIC{}, the LDQ DMs are typically in best agreement with experimental values for the absolute DM, and always respect the symmetries of the molecules in this study.
However, some LT DMs do not do so, leading to artifactual dipole moments as showcased above for azulene.

\subsubsection{Non-aromatic molecules}
\label{sec:nonaromatic_DIP16}

Both LT and LDQ describe identical chemical bonding patterns for non-aromatic parts of the molecules in this study; this means there is only one FOD geometry for such structures. 
Our test set contains 3 non-aromatic molecules which do not have any aromatic fragments: p-benzoquinone, p-chloranil, and p-fluoranil.
These molecules are mol04, mol07, and mol09 in our database; see \tabref{DIP16}.
They all have a \DTh{} molecular geometry; thereby, none should have a DM.
This is correctly reflected by all methods (DFT, RSIC, CSIC, FLO-SIC) used in this study.

For molecules which \emph{do} have aromatic fragments, like dichlone (mol08), fluorene (mol10), and 1,4-naphthalenedione (mol11), various electronic geometries can be generated for their aromatic part, allowing for a greater variety in the FOD search space. 
The resulting DMs point along the C$_{2}$ symmetry axis.
An exception is the LT3 structure of fluorene (mol10) that has a crooked DM with an angle of about 31$^{\circ}$ to the symmetry axis in \ERKALE{} and \PyFLOSIC{}.
This is unsurprising per the discussion in \secref{aromatic_DIP16}, as the corresponding electronic geometry shows 4 C-C bond FODs on one side of the inner C-ring, while there are 6 C-C bond FODs on the other side; this imbalance leads to a bias that tilts the electron density to one side of the molecule during the optimization.

As before, the DMs that arise from various LT and LDQ FOD arrangements turn out to be different in size.
DMs from LDQ FODs yield the best agreement of all SIC DMs with experimental reference values.

\subsubsection{Anti-aromatic molecules}
\label{sec:antiaromatic_DIP16}

The antiaromatic cyclobutadiene (mol15) molecule has two sets of C-C bond lengths:  1.43~\AA{} and 1.35~\AA{}.
According to its \DTh{} point group, the DM should be zero.
Similarly to \secref{benzene_results}, two LT and one LDQ structure can be formed.
Given the distinct C-C bond lengths, one can expect the LT structure in which the double-bond FODs are placed at the shorter C-C bonds to be energetically favorable. 
Our results are in line with these expectations, that is, the DM is zero in all calculations and the LT structures that place the double-bond FODs in the shorter C-C bonds are lower in energy than the structures that place the double-bond FODs on the longer C-C bonds.

\subsubsection{Summary}
\label{sec:fods_dipoles}

Having discussed the general features of the results on the DIP16 set of molecules, we can proceed with a summary analysis of the DIP16 data.
The mean error (ME)
\begin{equation}
    \text{ME} = \frac{1}{N} \sum_i \left( \mu_{i,\text{calc}} - \mu_{i,\text{ref}} \right)
    \label{eq:ME}
\end{equation}
and mean absolute error (MAE)
\begin{equation}
  \text{MAE} = \frac{1}{N} \sum_i \left| \mu_{i,\text{calc}} - \mu_{i,\text{ref}} \right|
  \label{eq:MAE}
\end{equation}
are calculated over the entire DIP16 test set, and are analyzed in the following.
The explicit values and  more details about the used reference values \cite{Mcclellan1963_1,Gelessus1995_505,Tobiason1973_773,Satoh2007_1103,El2011_1227,Dobyns1963_3553,Battaglia1981_421} can be found 
in the SI.

The PW92 DFT calculations yield a ME of $-0.06$~D and a MAE of $+0.10$~D both in \ERKALE{} and \PySCF{}; the excellent agreement again confirms the reproducibility and comparability of the results.
The MAEs for the SIC calculations from various starting points are shown in \tabref{table_dips_errors2}; the agreement with experiment is worse with SIC than without it, as the errors in \tabref{table_dips_errors2} are greater than the DFT values given above.
In \tabref{table_dips_errors2}, LT and LDQ stand for initialization with FODs/FLOs arising from these two methods; as some of the molecules afford many possible LT structures, the LT configuration leading to the lowest energy was chosen in each case.
An analogous procedure was also employed for LDQ.
Finally, $E_{\text{min}}$ refers to using the energetically lowest-lying solution, regardless of whether it represents LT or LDQ type bonding.

\begin{table}[h]
 \centering
 \caption{Mean error (ME) and mean absolute error (MAE) of absolute DMs for DIP16 in D for RSIC calculations in \ERKALE{} and FLO-SIC calculations in \PyFLOSIC{} initialized by various methods (see main text).} 
 \begin{tabular}{ ll | ccc }
                                &       & LT        & LDQ     & $E_{\text{min}}$ \\\hline
 \multirow{2}{*}{\ERKALE{}}     & ME    & $+0.22$   & $+0.19$ & $+0.22$          \\
                                & MAE   & $+0.29$   & $+0.25$ & $+0.28$          \\\hline
 \multirow{2}{*}{\PyFLOSIC{}}   & ME    & $+0.19$   & $-0.01$ & $+0.21$          \\
                                & MAE   & $+0.33$   & $+0.11$ & $+0.33$          \\                                                             
 \end{tabular}
 \label{tab:table_dips_errors2}
\end{table}

The RSIC calculations started from FODs in \ERKALE{} result in DMs that agree with the corresponding \PyFLOSIC{} values in most cases.
In cases where RSIC and FLO-SIC disagree, we found that the stability analysis and the subsequent re-optimization of the density in \ERKALE{} has taken the RSIC solution away from the FOD starting point; without the stability analysis, the agreement between RSIC and FLO-SIC is recovered.
This indicates that FLO-SIC does not correspond to a local minimum of RSIC in such cases.

Interestingly, the best DMs are obtained when starting from LDQ FODs, both in FLO-SIC calculations with \PyFLOSIC{} as well as RSIC calculations \ERKALE{}; in \PyFLOSIC{}, the MAE is close to the DFT value.
Worse DMs are obtained when analyzing DMs from FODs based on LT, or the overall energetically lowest solutions ($E_{\text{min}}$) which likewise results in the use of some LT structures instead of LDQ structures.

\subsubsection{Validation of computational methodology}
\label{sec:methodvalidation}

To emphasize that the results obtained with the PW92 functional and the pc-1 basis set are qualitatively correct, additional DMs of azulene (mol03), see \secref{aromatic_DIP16}, were computed using \PyFLOSIC{} with other exchange-correlation functionals and larger basis sets.
Starting out with the question of the exchange-correlation functional, additional calculations were performed with the PBEsol\cite{Perdew2008_136406} and r$^{2}$SCAN\cite{Furness2020_8208} functionals, which are GGA and meta-GGA approximations, respectively.
PBEsol has been identified as a particularly good functional for SIC calculations,\cite{Lehtola2016_4296} whereas r$^2$SCAN exemplifies the state of the art in non-empirical density functional approximations.
For these tests, the FODs were re-optimized for each functional.
As can be seen from the results in \tabref{table_mol03_XC}, the DMs from PBEsol are close to those from PW92, while r$^2$SCAN leads to larger dipole moments than PW92 or PBEsol. 

However, most importantly for the present study, the orientation of the DMs are similar in each case: the DMs based on LT FODs are noticeably crooked, confirming that the same issues can be found at other levels of theory and validating the findings of \citeref{Lehtola2016_3195} for the case of FLO-SIC.

\begin{table}[h]
 \centering
 \caption{DMs $\mu$ for azulene (mol03) using various exchange-correlation 
 functionals, as well as the angle $\alpha_\text{LT}$ between the DMs from 
 the two LT structures and the molecular symmetry axis, see \figref{erkale_mol03_dips}. 
 The pc-1 basis set was used in all calculations.}
 \begin{tabular}{ l | ccc }
                            & PW92          & PBEsol        & r$^{2}$SCAN   \\\hline 
 $\mu_{\text{DFT}}$         & 0.96~D        & 0.96~D        & 1.01~D        \\
 $\mu_{\text{LT}}$          & 0.68~D        & 0.66~D        & 0.97~D        \\
 $\mu_{\text{LDQ}}$         & 0.59~D        & 0.55~D        & 0.86~D        \\\hline
 $\alpha_\text{LT}$         & $10.1^\circ$  & $9.9^\circ$   & $6.2^\circ$   \\
 \end{tabular}
 \label{tab:table_mol03_XC}
\end{table}

Next, the basis question.
Repeating the calculations in the split-valence,  polarized double-$\zeta$, polarized triple-$\zeta$, and polarized quadruple-$\zeta$ pc-0, pc-1, pc-2, and pc-3 
basis sets, respectively, as well as their counterparts augmented with diffuse basis functions, the aug-pc-0, aug-pc-1, aug-pc-2, and aug-pc-3 basis sets, respectively, we obtain the results in \tabref{table_mol03_basis}.
As these calculations are computationally costly, and as we are only interested in the basis set convergence, the FODs were not reoptimized; instead, they were frozen to the pc-1 optimized values.

\begin{table}[h]
 \centering
 \caption{DMs $\mu$ and the angle $\alpha_\text{LT}$ between the LT DMs and the symmetry axis for azulene (mol03) using various pc-$n$ and aug-pc-$n$ basis sets and the PW92 functional.}  
 \begin{tabular}{ l | cccc } 
                            & pc-0          & pc-1          & pc-2          & pc-3 \\\hline
 $\mu_{\text{DFT}}$         & 1.13~D        & 0.96~D        & 0.91~D        & 0.88~D      \\
 $\mu_{\text{LT}}$          & 0.88~D        & 0.68~D        & 0.62~D        & 0.60~D      \\
 $\mu_{\text{LDQ}}$         & 0.82~D        & 0.59~D        & 0.52~D        & 0.50~D      \\\hline
 $\alpha_\text{LT}$         & 9.2$^\circ$   & 10.1$^\circ$  & 10.4$^\circ$  & 9.9$^\circ$ \\\
                            & aug-pc-0      & aug-pc-1      & aug-pc-2      & aug-pc-3 \\\hline
 $\mu_{\text{DFT}}$         & 1.09~D        & 0.94~D        & 0.88~D        & 0.88~D    \\
 $\mu_{\text{LT}}$          & 0.84~D        & 0.66~D        & 0.60~D        &        \\
 $\mu_{\text{LDQ}}$         & 0.78~D        & 0.57~D        & 0.51~D        &        \\\hline
 $\alpha_\text{LT}$         & 9.3$^\circ$   & 9.5$^\circ$   & 10.0$^\circ$  &        \\
 \end{tabular}
 \label{tab:table_mol03_basis}
\end{table}

The DFT DMs arising from the calculations in the aug-pc-2, pc-3, and aug-pc-3 basis sets are identical, meaning that the DM has converged to the complete basis set limit.
Similarly, there is excellent agreement in the SIC DMs between the pc-3 and aug-pc-2 calculations; SIC calculations were not attempted in aug-pc-3 due to the large size of this basis set.
We can thus confirm that the data with the augmented triple-$\zeta$ aug-pc-2 basis set is (almost) at the complete basis set limit.

Having established the basis set limit, we can comment on the implications of the basis set converge on the present study.
Although improving the basis affects the absolute values of the DMs, the angle between the LT DMs to the symmetry axis remains at roughly 10$^{\circ}$.
This illustrates that the symmetry-broken solutions are not artifacts of small basis sets and the results obtained with the pc-1 basis are reliable.

The sufficiency of the pc-1 basis for the present study is further supported by PW92 DFT calculations for the whole DIP16 set in the aug-pc-2 basis.
The ME and MAE values for pc-1 are in excellent agreement with values from the aug-pc-2 calculations, which are $-0.05$~D and 0.10~D, respectively.
Since the DFT dipole moments coincide with the molecular plane, the importance of diffuse functions is less pronounced, as the molecular plane already has a decent amount of flexibility from the valence basis functions in the pc-1 basis set.

As a note, we tested whether fixing the FODs  corresponding to the 1s orbitals to the nuclear sites helps to converge the FLO-SIC calculations without compromising the resulting DMs.
As can be seen from the results presented in the SI, the DMs resulting from such a procedure are in excellent agreement with the DMs from the fully optimized FODs.
Although this strategy was not exploited in this work, it could be used in future investigations to achieve better computational efficiency.

\subsection{Further investigations on bonding motifs}
\label{sec:further}

We would like to point out that while we have investigated several types of FOD configurations for the DIP16 database, we have not made any claims to have found the best possible FOD configurations for any of the systems examined in this work.
Even though we have examined several kinds of electronic geometries in our calculations, that is, multiple LT and LDQ structures, it is possible that some other type of FOD configuration yields even lower energies.
However, all the FOD configurations employed in this work are fully reproducible and can even be generated automatically, thus delivering reproducible FLO-SIC solutions; the molecular and FOD configurations used in this work have also been included in the SI.

For the case of benzene, fascinating bonding motifs in which all three double-bond FOD pairs hover out-of-plane on the same side of the molecule have been found and described by \citet{Pederson2015_153} for a symmetric molecular geometry for benzene; correspondingly, we will term these \emph{hovering FODs}.
It is likely that several of the molecules in the DIP16 database likewise afford hovering FOD configurations, given that they are cyclic, planar molecules like benzene.
However, the role of the hovering bonding motifs is still not clearly understood; at present, we also lack the tools to generate them in an automatic manner.
Until now, despite thorough efforts, we have only found stable hovering FODs for ring systems.
In contrast, hovering double-bond FOD pairs are not suitable solutions for ethene, (C$_{2}$H$_{4}$), for example, as such FODs converge back to a LT solution upon optimization.

The DIP16 database contains molecules that were also considered by \citet{Adhikari2020_184303}.
\citeref{Adhikari2020_184303} describes using fixed molecular geometries from a chemical database, and optimizing only a single FOD configuration for each molecule.
Because of the problem of many local minima first pointed out in \citeref{Lehtola2016_3195} for variational SIC calculations and extended to FLO-SIC in this work, it is unlikely that such a procedure leads to the global minimum for FLO-SIC.
Comparing our FOD configurations with the ones of \citeauthor{Adhikari2020_184303}, obtained directly from the authors,\cite{Adhikari2021_0} we observed that their FOD configurations place some double-bond FOD pairs above the molecular plane for acridine (mol01) and phenazine (mol12); the FOD structures for these molecules are thus partly hovering in nature.

\begin{figure}
\centering
\subfigure[1UH]{\includegraphics[width=0.155\textwidth]{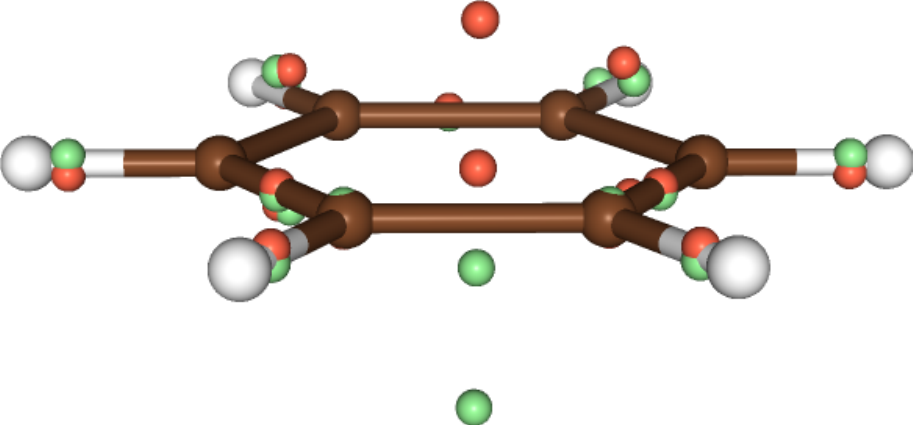}} \qquad \subfigure[2UH]{\includegraphics[width=0.155\textwidth]{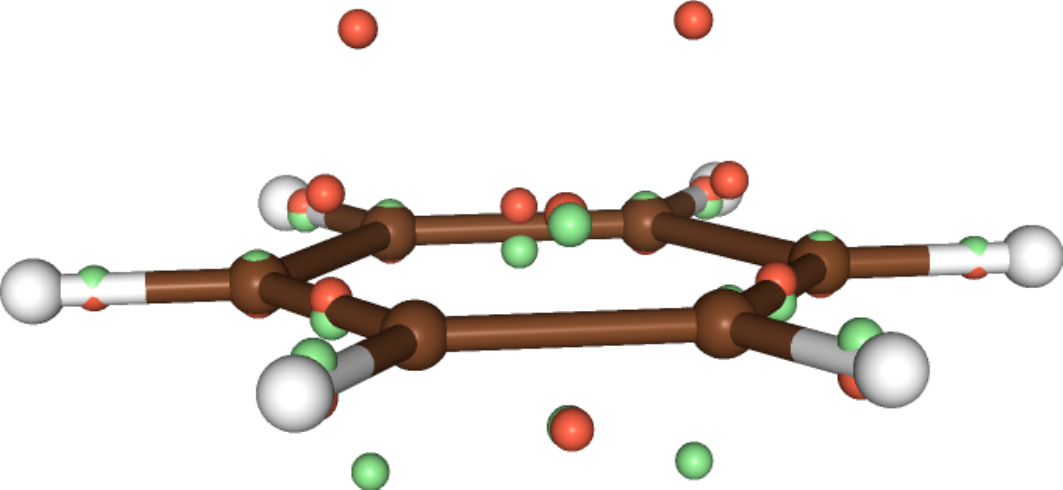}} \qquad \subfigure[3UH]{\includegraphics[width=0.155\textwidth]{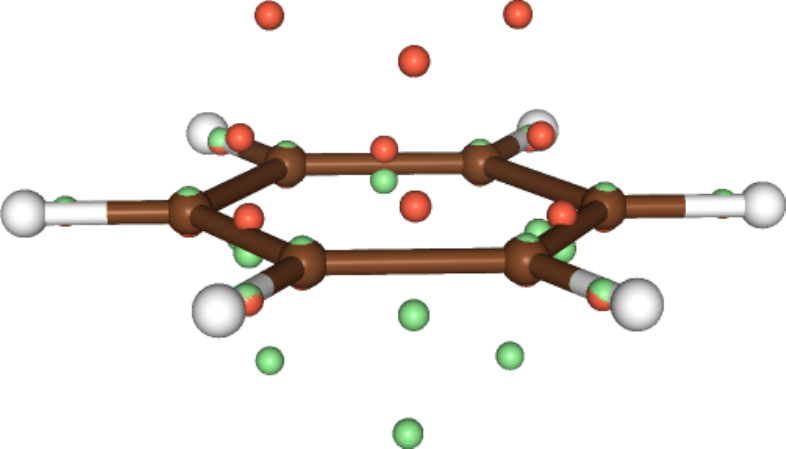}} \qquad \subfigure[3RH]{\includegraphics[width=0.155\textwidth]{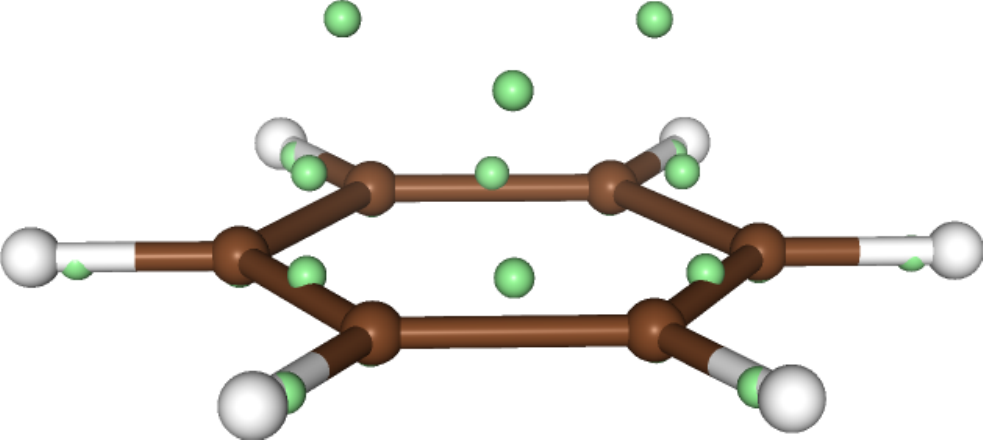}}
 \caption{Examples of optimized hovering FODs in benzene. UH stands for unrestricted hovering, RH stands for restricted hovering, and the numbers denotes the number of double bonds hovered. Note that all other FODs are affected by hovering any number of double bonds.} 
 \label{fig:figure_benzene_hovering}
\end{figure}

In order to try to understand such solutions, 
we return to our favorite model system---benzene---and start out by examining the FOD structure described by \citet{Pederson2015_153}.
Employing the methodology described in \secref{methods} to optimize this configuration, denoted as 3RH in \figref{figure_benzene_hovering}, we obtain a large out-of-plane dipole moment (ca. 0.10~D) and an energy of $-233.013171~E_{\text{h}}$, which is lower than that of our LT solution ($-233.007512~E_\text{h}$).

Obviously, such a dipole moment is completely artifactual, and is at variance to experiment and results of \emph{ab initio} calculations.
The artifactual dipole moment can be eliminated for this molecule by splitting the $\alpha$ and $\beta$ FODs onto different sides of the molecule; we denote this novel FOD pattern as unrestricted hovering; it is illustrated as 3UH in \figref{figure_benzene_hovering}.
However, as the spin symmetry is now broken, this procedure leads to a slightly non-zero spin density and an $\langle \hat{\bf S}^2 \rangle \approx 5 \times 10^{-4}$.

However, also other kinds of hovering solutions are possible: instead of hovering all three double-bond FOD pairs, one can also hover just one or two of them, and fully relax the FODs in an unrestricted calculation; the structures are presented in \figref{figure_benzene_hovering}.
Hovering one pair lowers the energy (1UH, $E=-233.009569~E_\text{h}$), but induces a large dipole moment (0.28~D) in the molecular plane. 
Next, hovering two pairs lowers the energy further (2UH, $E=-233.011030~E_\text{h}$), still increasing the dipole moment (0.29~D).        
Only when all three pairs are hovered is the correct dipole moment of zero obtained along with an even lower energy (3UH, $E=-233.0132249~E_\text{h}$).

Next, starting from the lattermost structure, in which the three hovering double-bond FOD pairs form a triangle, one can vary the the size of the triangle, as well as the distance of the triangle from the bonding plane.
For an unrestricted calculation, the $\alpha$ and $\beta$ triangles' sizes and distance from the molecular plane can be varied to obtain even more hovering structures.
The currently lowest energy we have found ($-233.015872~E_\text{h}$) is produced by a hovering structure where the hovering $\alpha$ and $\beta$ triangles are rotated against each other by $60^{\circ}$; this motif thereby belongs to the LDQ class of motifs.

As the above discussion demonstrates, determining the lowest possible solution is non-trivial even for benzene.
Larger molecules as those included in the DIP16 set may afford several types of solutions with hovering FODs, and finding these solutions requires global optimization techniques due to the problem of multiple local minima: the two hovering structures employed by \citet{Adhikari2020_184303} can likely be further tuned to give a significantly lower energy by hovering more or all of the double-bond FOD pairs. 

All FLOs corresponding to hovering FODs appear to be less localized than FLOs from the respective LT or LDQ FODs, both according to visual inspection, as well as according to the Foster--Boys localization criterion.
This is another aspect which needs to be studied in more detail in the future. 
More work is thus needed for a full understanding of any potential hovering FOD solutions.

Moreover, it is important to note that each electronic geometry is likely to correspond to a different molecular geometry.\cite{Lehtola2016_3195}
When the geometry is optimized for the LT structure, yielding a symmetry-broken molecular geometry with alternating single and double bonds,\cite{Lehtola2016_3195} the hovering FOD solutions no longer yield a decrease in energy.

\section{Summary and discussion}
\label{sec:summary}

As originally pointed out in \citeref{Lehtola2016_3195}, symmetry breaking is an important problem in self-interaction corrected calculations. 
We have shown that Fermi-orbital descriptors (FODs) and the corresponding Fermi--L\"owdin orbitals (FLOs) can be used to guide self-interaction corrected calculations towards the wanted type of solution, that is, the wanted type of local minimum.
In the first part of the manuscript we showed that symmetry-broken solutions for benzene in real (RSIC) and complex orbital SIC (CSIC) can be reproduced starting from Lewis-like solutions.
We demonstrated that a symmetric molecular geometry for benzene can be captured with SIC by starting the calculation from a Linnett double-quartet (LDQ) structure in RSIC, CSIC, and in FLO-SIC.
While the symmetric LDQ solution is higher in energy than the Lewis structures when the molecular geometry is allowed to relax, the LDQ structure is the lowest-lying solution for the symmetric molecular geometries considered in this study.
Moreover, the millihartree energy difference between Lewis' theory (LT) and LDQ structures is negligible with respect to the overall SIC contribution which is in the order of hartrees. 

In the second part of this study, a benchmark set consisting of 16 planar, cyclic molecules we call DIP16 was used to show that Lewis-like SIC solutions often break the symmetry of the electronic density,
while LDQ SIC solutions do not.
Reasonable LDQ solutions can typically be constructed by combining two Lewis-like solutions, by taking the spin-up FODs from one Lewis structure and the spin-down FODs from another Lewis structure.
Thus, we were able to demonstrate that LDQ SIC is able to provide a simple representation of resonance/aromaticity.

\begin{figure}
    \centering
    \includegraphics[width=0.49\textwidth]{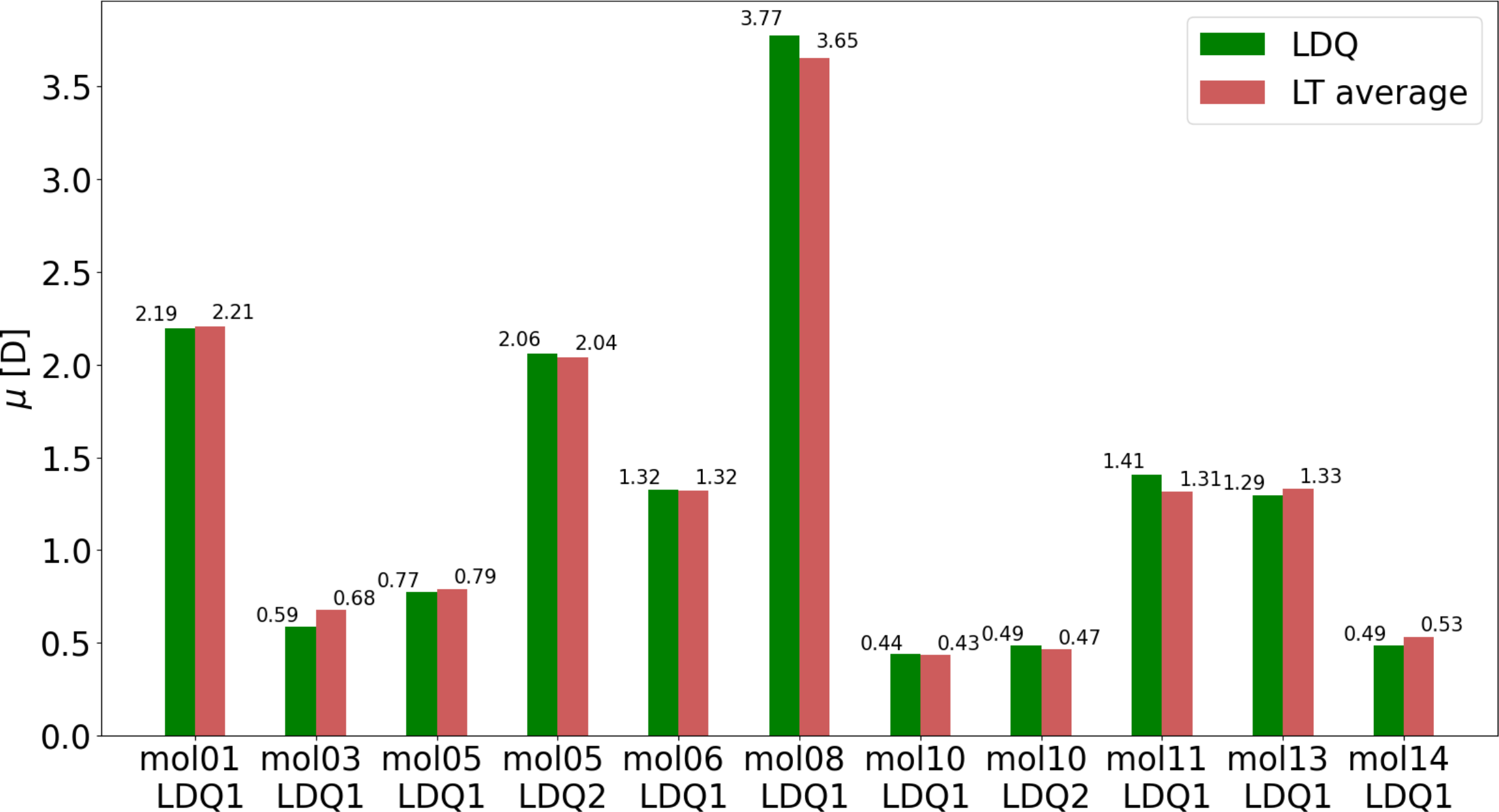}
     \caption{DIP16: Comparison of  LDQ DMs and the corresponding averages of LT DMs  from  \PyFLOSIC{} calculations; the DMs as well as the averaging procedure can be found in the SI.
     See \tabref{DIP16} for the legend of molecules.}
    \label{fig:comp_resonance}
\end{figure}

Interestingly, when the resulting Linnett structure is optimized, a dipole moment that coincides roughly with the average of the dipole moments resulting from the two employed Lewis structures is obtained.
To quantify this, for our benchmark set a linear combination of LT DMs is compared to the corresponding LDQ DM  in \figref{comp_resonance}; the resulting DMs agree well. 
It shall be noted that there are cases which require more than two Lewis structures to obtain the LDQ structure, which is the case for acridine and fluorene in the DIP16 set (see the SI).

In summary, we have pointed out that FODs guided by Lewis' theory of chemical bonding tend to result in symmetry breaking, while FODs based on Linnett's theory are able to restore qualitatively correct total densities for the systems studied, at the cost of breaking the spin symmetry.

Molecular symmetries appear to break whenever the FOD arrangement does not have the same symmetry as the molecule.
Classifying the FODs in a given molecule with respect to its symmetry-equivalent parts, given by mirror planes or rotations, for example, one finds that if the number of FODs for the symmetry-equivalent parts of the molecule differ, the DM from a self-consistent calculation will point towards the part of the molecule that has more FODs; that is, the density follows the FOD arrangement. 
Any imbalances in the FOD arrangement result in a bias for the electron density in SIC calculations, as the SIC prefers localized orbitals that are determined by the FODs.

Our findings are not limited to the FLO-SIC method, as we were also able to reproduce the main findings of this work also for RSIC and CSIC by starting the calculations from FLOs guided by LT or LDQ FODs.

A careful initialization of SIC calculations is of paramount importance.
For instance, various starting points may lead to convergence to different local minima.
In many cases, several FOD arrangements have to be evaluated to find the minimum-energy SIC solution.
Employing a single type of FOD arrangement may lead to incorrect conclusions.

Concluding the summary, we move on to further discussion.
The second part of this study---the examination of the DIP16 dataset described in this work---employed fixed molecular geometries.
Optimizing the geometries will likely bias the dipole moments further if the FODs break 
the molecular symmetry.
New advances in theory are warranted in order to improve the usefulness of self-interaction corrected calculations for practical applications, which typically require the determination of optimal molecular geometries, for instance, and we hope to address such topics in future work.

Furthermore, we excluded hovering FOD structures from the current investigations on the DIP16 dataset.
Although all presently-known hovering structures can be constructed starting from a LT or LDQ configuration by moving some of the FODs out-of-plane, there is currently no elegant, systematic way to set up reproducible hovering motifs for more complex molecules, such as those included in the DIP16 benchmark set of this work. 
The phenomenon of hovering FODs is not yet understood, and we hope to find suitable explanations in further research. 

As shown in this study, several kinds of FOD configurations and FLO-SIC solutions are possible for molecules as simple as benzene, as could be expected from the results of  \citeref{Lehtola2016_3195} for fully variational SIC.
We have demonstrated in \secref{methods} that the various kinds of FLO-SIC solutions all correspond to local minima at symmetric nuclear geometries for benzene.
Importantly, as shown here and in \citeref{Lehtola2016_3195}, if the nuclear geometry is relaxed, the resulting optimal molecular geometries may be different for each type of solution, further complicating the problem of locating the lowest-lying solution.

Finally, we would also like to point out that free and open-source software is essential for a fruitful development of 
computational methods,\cite{Oliveira2020_024117, Lehtola2021__} as they enable straightforward one-to-one comparisons as in this work. 
All authors of this paper are either main developers of, or 
contributors to various free and open-source software electronic structure programs, making self-interaction correction methods readily available to a broad community of physicists,  chemists, and material scientists.

\section*{Acknowledgment}
We thank all the sponsors and guests of the "Quo Vadis
Self-Interaction Correction?" workshop (Freiberg, Germany, 2019), as this meeting brought some of the present authors together in person for the first time and thereby enabled them to start a fruitful cooperation.
Moreover, it was an honor to meet Prof. John P. Perdew, the heart and one of the most important persons in SIC and a living legend in the field of electronic structure.
Further thanks goes to Prof. Mark R. Pederson for creating the FLO-SIC method and Prof. Hannes J\'onsson for major contributions in the development of the CSIC method,
which are both methods discussed within the manuscript. 
This contribution is part of our OpenSIC project, which aims to develop fresh perspectives in the field of self-interaction corrections using open-source software.
J. Kortus and S. Schwalbe have been funded by the Deutsche
Forschungsgemeinschaft (DFG, German Research Foundation) - Project ID 421663657 - KO 1924/9-1.
We acknowledge the contributions of Jakob Kraus at an early stage of this project.
We furthermore thank the ZIH in Dresden for computational time and support. 

\section*{Supporting information}
See the supporting information for thorough information on the used molecular geometries, the full data set of dipole moments, comparison of the LDQ dipole moments as averages of the respective LT dipole moments, the molecule-by-molecule discussion of the electronic geometries and the resulting dipole moments of the DIP16 data set, as well as the analysis of freezing the $1s$ FODs at the nuclei.
The full set of molecular and electronic geometries is also available as part of the supporting information.

\attachfile[color=0 1 0]{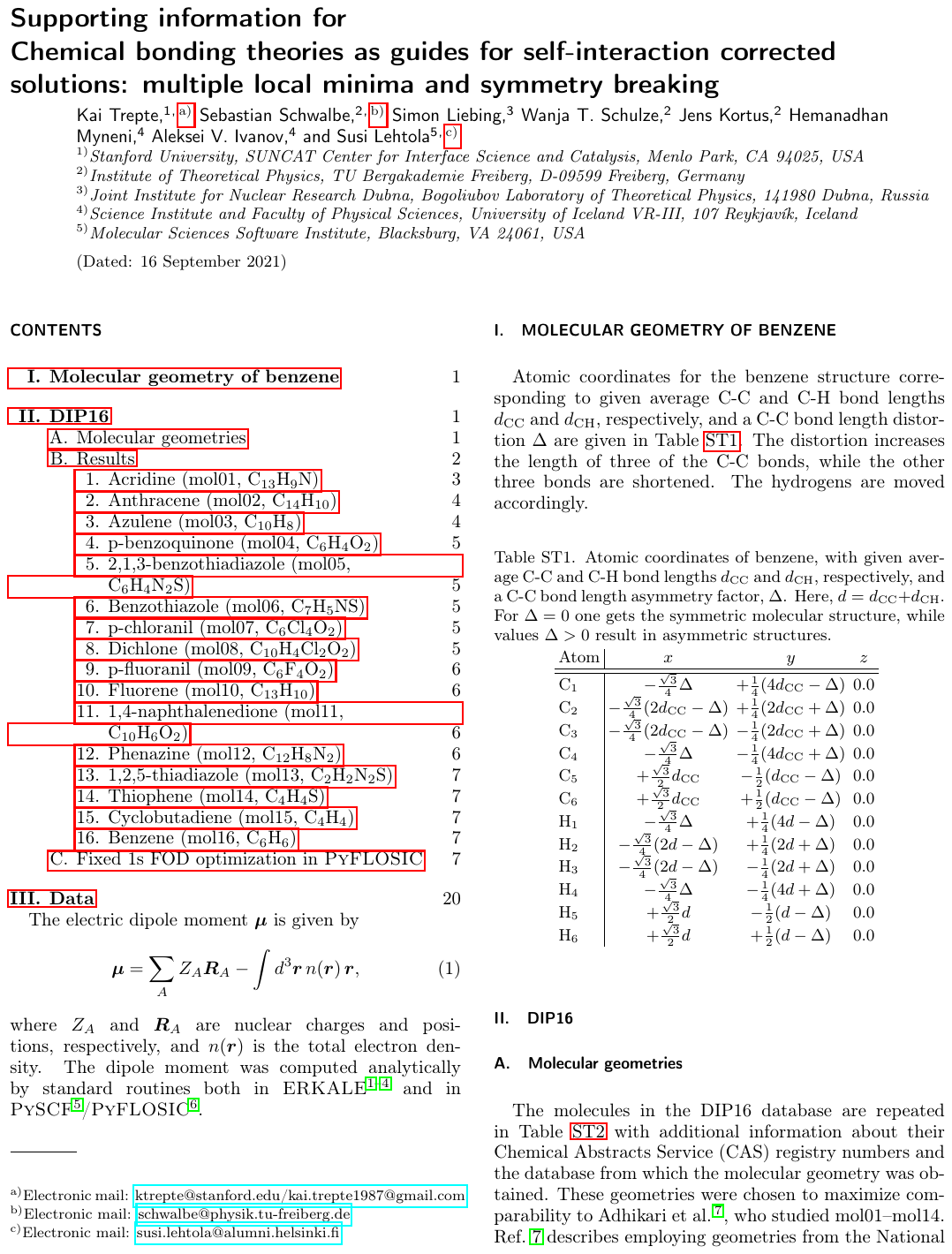}{Click on the pin for the supporting infomation.}

\section*{Data Availability Statements}
The data that supports the findings of this study are available within the article and its supporting information.
The data are also openly available in \url{https://gitlab.com/kaitrepte/dip16}.

\section*{Author Declarations}
The authors have no conflicts to disclose.

\bibliography{ref.bib}
\end{document}